\documentclass[11pt,onecolumn,a4paper]{article}

\usepackage{fullpage}
\usepackage{amsfonts}
\usepackage{amsthm}
\usepackage{amsmath}
\usepackage{verbatim}
\usepackage[utf8]{inputenc}
\usepackage{float,tikz}
\usetikzlibrary{decorations.pathreplacing}
\usetikzlibrary{decorations.markings}
\usetikzlibrary{shapes}

\tikzset{
  spun/.style={line width=4pt,gray}
}
\newcommand\drawvert[2]{\fill (#1,#2) circle (0.1cm)}
\newcommand\drawvertb[2]{\fill (#1,#2) circle (0.3cm)}

\newcommand\Qnonneg{\mathbb{Q}^{\geq 0}}
\newcommand\eee[1]{{\mathbf{e}_{#1}}}

\newcommand\Prat{\theta}
\newcommand\dist[2]{\operatorname{d}(#1,#2)}

\newcommand\defn[1]{{\bf #1}}

\newcounter{all}
\newtheorem{theorem}[all]{Theorem}

\newtheorem{lemma}[all]{Lemma}
\newtheorem{proposition}[all]{Proposition}
\newtheorem{definition}[all]{Definition}
\newtheorem{remark}[all]{Remark}

\newcommand{\sem}[1]{\left[\!\left[#1\right]\!\right]}
\newcommand\cp[1]{\ensuremath{\mathsf{\##1}}}
\newcommand\napp{\cp{ParityNAE}}
\newcommand\nCSP{\cp{CSP}}
\newcommand\nSFO{\cp{SFO}}
\newcommand\nPM{\cp{PerfMatch}}
\newcommand\nSAT{\cp{SAT}}

\newcommand\FugacityWeightedPM{\cp{FugacityWeightedPM}}
\newcommand\nP{\cp{P}}
\newcommand\calF{\mathcal F}

\newcommand\vecphi{\boldsymbol {\phi}}
\newcommand\veczero{\boldsymbol {0}}
\newcommand\vecx{\mathbf x}
\newcommand\vecy{\mathbf y}
\newcommand\vecz{\mathbf z}
\newcommand\vecw{\mathbf w}
\newcommand\vecp{\mathbf p}
\newcommand\wt{\operatorname{wt}}
\newcommand\APred{\leq_\mathrm{AP}}
\newcommand\sig{weight-function}
\newcommand\sigs{weight-functions}

\newcommand\sigf[1]{:\{0,1\}^{#1}\to\Qnonneg}
\newcommand\Holant{\operatorname{Holant}}

\newcommand\Match[1]{\mathcal{M}_{#1}}
\newcommand\Matchp[1]{\mathcal{M}'_{#1}}

\newcommand\relEven{\mathrm{Even}}
\newcommand\relOdd{\mathrm{Odd}}
\newcommand\relOR{\mathrm{OR}}
\newcommand\relNAE{\mathrm{NAE}}
\newcommand\relEvenNAE{\mathrm{EvenNAE}}
\newcommand\relEvenOR{\mathrm{EvenOR}}
\newcommand\sigEven{\mathbf{Even}}
\newcommand\sigOdd{\mathbf{Odd}}
\newcommand\sigOR{\mathbf{OR}}
\newcommand\sigNAE{\mathbf{NAE}}
\newcommand\sigEvenNAE{\mathbf{EvenNAE}}
\newcommand\sigEvenOR{\mathbf{EvenOR}}
\newcommand\Edge{\mathrm{Edge}}
\newcommand\Fugacity{\mathrm{Fugacity}}

\newcommand\drawinc[2] {\filldraw[fill=white] (#1,#2)+(-0.1,-0.1) rectangle +(0.1,0.1);}
\newcommand\drawincg[2] {\filldraw[fill=black] (#1,#2)+(-0.1,-0.1) rectangle +(0.1,0.1);}

\tikzstyle{bigv}=[circle, draw]

\makeatletter
\newcommand\prob[3]{\goodbreak\begin{list}{}{\labelwidth\z@ \itemindent-\leftmargin
                        \itemsep\z@  \topsep6\p@\@plus6\p@
                        \let\makelabel\descriptionlabel}
                \item[\it Name]#1
                \item[\it Instance]#2
                \item[\it Output]#3
                \end{list}}
\makeatother

\makeatletter
\newtheorem*{rep@theorem}{\rep@title}
\newcommand{\newreptheorem}[2]{%
\newenvironment{rep#1}[1]{%
 \newcommand\rep@title{#2 \ref{##1}}%
 \begin{rep@theorem}}%
 {\end{rep@theorem}}}
\makeatother

\newreptheorem{corollary}{Corollary}
\newreptheorem{proposition}{Proposition}
\newreptheorem{theorem}{Theorem}
\newreptheorem{lemma}{Lemma}

\title{Approximating Holant problems by winding}
\author{Colin McQuillan\\
Ashton Building\\
Department of Computer Science\\
University of Liverpool\\
Liverpool L69 3BX\\
  \texttt{cmcq@liv.ac.uk}}
\date{\today}

\begin{document}

\maketitle

\abstract We give an FPRAS for Holant problems with parity constraints
and not-all-equal constraints, a generalisation of the problem of
counting sink-free-orientations.  The approach combines a sampler for
near-assignments of ``windable'' functions -- using the
cycle-unwinding canonical paths technique of Jerrum and Sinclair --
with a bound on the weight of near-assignments.  The proof generalises
to a larger class of Holant problems; we characterise this class and
show that it cannot be extended by expressibility reductions.

We then ask whether windability is equivalent to expressibility by
matchings circuits (an analogue of matchgates), and give a positive
answer for functions of arity three.

\section{Introduction}

In this paper we will show that the following problem has an FPRAS
(a type of approximation algorithm - see Section \ref{sec:compdef}).

\prob{\napp}
{A multigraph $G$ in which each vertex is labelled Even, Odd, or NAE}
{The number of subsets $F\subseteq E(G)$ such that:
\begin{itemize}
\item each Even vertex has an even number of incident edges in $F$
\item each Odd vertex has an odd number of incident edges in $F$
\item each NAE vertex has at least one incident edge in $F$ and at least one incident edge in $E(G)\setminus F$
\end{itemize}
}

\begin{theorem}\label{thm:nappalg}
There is an FPRAS for $\napp$.
\end{theorem}

\subsection{Relationships with other counting problems}

A \emph{sink-free orientation} of a graph is a choice of orientation
of each edge such that no vertex has out-degree zero.  The problem
$\nSFO$ is: given a graph, count the number of sink-free orientations.
We can also allow ``skew'' edges, where the ends of the edge must both
be oriented outwards or both oriented inwards.

\begin{figure}
  \centering
\begin{tikzpicture}[scale=2,decoration={
    markings,
    mark=at position 0.5 with {\arrow[scale=4]{>}}}
    ] 
    \draw[postaction={decorate}] (0,0)--(1,2);
    \draw[postaction={decorate}] (0,0)--(2,0);
    \draw[postaction={decorate}] (2,0)--(1.5,1);
    \draw[postaction={decorate}] (1,2)--(1.5,1);

\filldraw (0,0) circle (0.1cm)
  (1,2) circle (0.1cm)
  (2,0) circle (0.1cm);

\node at (3,1) {$\longleftrightarrow$};  

\begin{scope}[xshift=4cm,nodes={draw,fill=white}]
\draw (0,0)--(1,2)--(2,0)--(0,0);
\draw[line width=0.4cm, gray] (0.5,1) -- (1,2) (1,0) -- (2,0);
\draw[line width=0.4cm, gray] (0,0)--(-0.5,0.5)
 (1,2)--(1,2.5)
 (2,0)--(2.5,0.5);
\node[ellipse] at (0,0) {NAE};
\node[ellipse] at (1,2) {NAE};
\node[ellipse] at (2,0) {NAE};
\node[ellipse] at (0.5,1) {Odd};
\node[ellipse] at (1,0) {Odd};
\node[ellipse] at (-0.5,0.5) {Odd};
\node[ellipse] at (1,2.5) {Odd};
\node[ellipse] at (2.5,0.5) {Odd};
\end{scope}
\end{tikzpicture}
\caption{Reduction from $\nSFO$ to $\napp$. The edge with two
  arrows is a skew edge.  A sink-free orientation is illustrated with
  the corresponding set $F$ draw in thick grey.}\label{fig:sfo}
\end{figure}
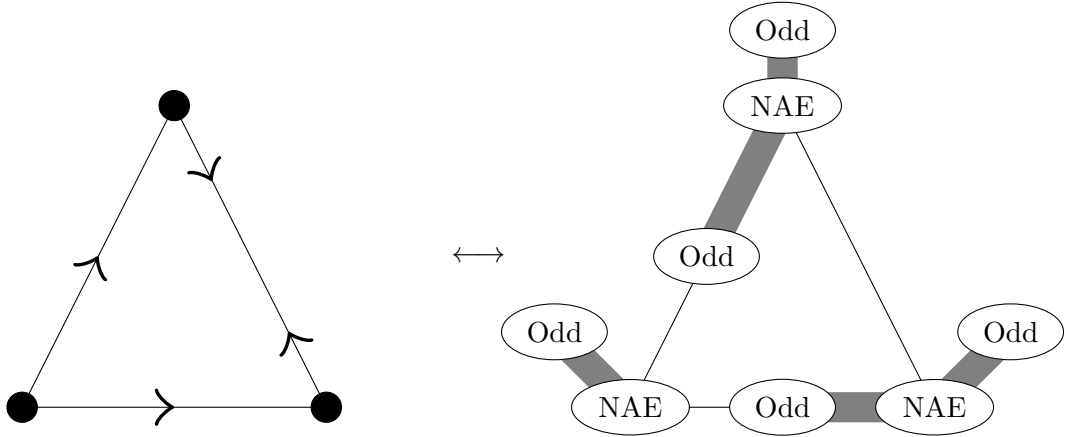

Bubley and Dyer studied $\nSFO$ and gave an FPRAS
\cite{bubleydyer}. They showed as a corollary that there is an FPRAS
for counting solutions to a formula in conjunctive normal form in
which every variable appears at most twice, which they showed is a
\#P-hard problem.  The first part of their argument was a standard
reduction to sampling - finding a fully polynomial almost uniform
sampler (FPAUS) for sink-free orientations.  Then, they constructed a
Markov chain that converges to the uniform distribution on sink-free
orientations, and bounded its mixing time using a two-stage path
coupling argument.  Cohn, Pemantle and Propp later gave an exact
sampler with $O(|V|\cdot |E|)$ mean running time using a kind of
rejection sampling \cite{cohn}.

A simple reduction from $\nSFO$ to $\napp$ is illustrated in Figure
\ref{fig:sfo}, showing that $\napp$ generalises the problem of
counting sink-free orientations in a graph (while also allowing parity
constraints).  Given an instance $G$ of $\nSFO$, label all the vertices
NAE, subdivide each non-skew edge $uv$, label the new vertices (which
we will refer to as ``$m_{uv}$'') Odd, then attach a degree-one Odd
vertex to each NAE vertex. This gives an instance $G'$ of $\napp$.
For all orientations $O$ of $G$ define a set $F_O\subseteq E(G')$ by
taking all edges attached to degree-one Odd vertices and all
``heads'': for non-skew edge $uv$ take $um_{uv}\in F$ if and only if
$uv$ is directed towards $u$, and for skew edges $uv$ take $uv\in F$
if and only if $uv$ is oriented outwards.  Each degree-two Odd vertex
in $G'$ has exactly one incident edge in $F_O$, and each NAE vertex in
$F_O$ has at least one incident edge in $F$, and if $O$ is sink-free
then each NAE vertex in $F_O$ has at least one incident edge not in
$F$.  Furthermore, any $F\subseteq E(G')$ satisfying these conditions
is $F_O$ for some sink-free orientation $O$.  The function $O\mapsto
F_O$ therefore gives a bijection from sink-free orientations of $G$ to
the set of subsets of $E(G')$ that get counted by $\napp$.

$\napp$, at least when restricted to bounded-degree graphs, is a type
of Boolean \emph{Holant problem}.  Holant problems are a quite general
type of graphical counting problem.  The constraints Odd, Even and NAE
in $\napp$ are generalised to functions $F\colon\{0,1\}^k\to\mathbb
C$, called (Boolean) signatures in this context.  Relations
$R\subseteq\{0,1\}^k$ can also be used by considering the function
$\mathbf R\colon\{0,1\}^k\to\{0,1\}$ that takes the value $1$ exactly on elements of
$R$.  In this discussion we will take the codomain to be the set of
complex numbers, but afterwards we will restrict to non-negative
rational-valued signatures, which we call \emph{weight-functions}.

A Holant instance is a graph $G$ equipped with a function $F_v\colon
\{0,1\}^{J_v}\to\mathbb C$ for each vertex $v$, where $J_v$ is the set
of edges incident to $v$.  In fact we will want to allow self-loops,
which calls for a slightly more complicated definition - see Section
\ref{sec:prelim}.  We are interested in the total weight
$$\sum_{\vecx\in\{0,1\}^{E(G)}}\prod_{v\in V(G)}F_v(\vecx|_{J_v}).$$
For example, if all $F_v$ take the value $1$ on vectors of Hamming
weight $1$, and take the value $0$ otherwise, then the total weight is
just the number of perfect matchings of $G$, because a vector
$\vecx\in\{0,1\}^{E(G)}$ is the characteristic vector of a perfect
matching of $G$ if and only if $\prod_{v\in V(G)}F_v(\vecx|_{J_v})=1$.

Given a finite set $\calF$ of signatures, $\Holant(\calF)$ is the
problem of evaluating the total weight where $G$ is given as input,
and where we require that each $F_v$ is a copy of some $F\in\calF$:
for some enumeration $v_1,\dots,v_k$ of $J_v$ we have
$F_v(\vecx)=F(\vecx(v_1),\dots,\vecx(v_k))$ for all
$\vecx\in\{0,1\}^{J_v}$.  For example, let $F$ be the function defined
by $F(0,0,1)=F(0,1,0)=F(1,0,0)=1$ and $F(i,j,k)=0$ elsewhere.  Then
$\Holant(\{F\})$ is the problem of counting perfect matchings in
degree-three graphs.

For all positive integers $k$ define
$\sigEven_k,\sigOdd_k,\sigNAE_k\colon\{0,1\}^k\to\{0,1\}$ by setting
$\sigEven_k(x_1,\dots,x_k)$ to be $1$ if and only if $x_1+\dots+x_k$ is even,
setting $\sigOdd_k(x_1,\dots,x_k)$ to be $1$ if and only if
$x_1+\dots+x_k$ is odd, and setting $\sigNAE_k(x_1,\dots,x_k)$ to be
$1$ if and only if $1\leq x_1+\dots+x_k\leq k-1$.  The restriction of
$\napp$ to graphs of maximum degree at most $d$ is then equivalent to
$\Holant(\calF_d)$ where
$\calF_d=\{\sigEven_1,\sigOdd_1,\sigNAE_1,\dots,\sigEven_d,\sigOdd_d,\sigNAE_d\}$.
By Theorem \ref{thm:nappalg}, this problem has an FPRAS for each $d$.

$\napp$ is also a counting constraint satisfaction problem,
at least when restricted to bounded-degree graphs.  Roughly
speaking, a instance of a constraint satisfaction problem is a list of constraints like ``$x\vee y$, $y\wedge z$, $y=z$'', and we are
interested in the number of configurations satisfying the constraints.
In particular \cite{lsm}, for any finite set of signatures $\calF$,
an instance of $\nCSP(\calF)$ is a set of variables $V$ and a list of
formal function applications
$$F_1(v_{1,1},\dots,v_{1,k_1}),\dots,F_s(v_{s,1},\dots,v_{s,k_s})$$
where $F_i\colon\{0,1\}^{k_i}\to\mathbb C$ is a function in $\calF$
for each $1\leq i\leq s$, and $v_{i,j}\in V$ is a variable for each
$1\leq i\leq s$ and $1\leq j\leq k_i$.  The value of this instance is
$$\sum_{\vecx\in\{0,1\}^V}\prod_{i=1}^m
F_i(\vecx(v_{i,1}),\dots,\vecx(v_{i,k_i})).$$
For information about the complexity of approximately evaluating $\nCSP$s,
 see \cite{bigdomain}.

A $\nCSP(\calF)$ instance can be drawn as a ``dual constraint
hypergraph'', which has vertices for each of the constraints
$c_1,\dots,c_s$, and a hyperedge $\{c_i\mid v=x_{i,j}\}$ for each
variable $v$ (ignoring multiplicities for now).  The dual constraint
hypergraph is a graph if and only if every variable appears exactly
twice. In this way, $\Holant(\calF)$ is the restriction of
$\nCSP(\calF)$ to read-twice instances.  Note that the variables of
the $\nCSP$ are the edges of the Holant instance; sometimes a $\nCSP$
is described in the opposite way, as the primal constraint hypergraph,
with variables as vertices and $s$ edges or hyperedges.

We now recall the relationships between $\nCSP$s and Holant problems in
the Hadamard basis as discussed in \cite{planarcsp}.  Note that while
these equivalences are usually stated in the context of exact
evaluation, the reductions just involve preprocessing the input, and
so also apply in the context of approximate counting.  Firstly,
equality constraints can be used to break the read-twice
restriction. Letting $=_3$ denote the function $\{0,1\}^3\to\mathbb C$
taking the value $1$ on $(0,0,0)$ and $(1,1,1)$, and taking the value
$0$ elsewhere, if $=_3$ is in $\calF$ then $\Holant(\calF)$ is
equivalent to $\nCSP(\calF)$ \cite[Proposition 5]{holant_and_back}.
Secondly, let $\widehat F\colon\{0,1\}^k\to\mathbb C$ denote the
Hadamard transform, defined by
$$\widehat
F(x_1,\dots,x_k)=2^{-k/2}\sum_{\vecy\in\{0,1\}^k}F(y_1,\dots,y_k)(-1)^{x_1y_1+\dots+x_ky_k}.$$
$\Holant(\calF)$ is always equivalent to $\Holant(\{\widehat F \mid
F\in\calF\})$; see \cite[Proposition 1]{holant_and_back} or
\cite{Mao}.  Also, $\widehat {\widehat F}=F$ for any $F$. So if
$\calF$ contains $\widehat{=_3}$, then $\Holant(\calF)$ is equivalent
to $\nCSP(\{\widehat F\mid F\in \calF\})$.  But $\widehat{=_3}$ is
just $\sigEven_3$ multiplied by a factor of $\sqrt{2}$ (which can be
easily accounted for).

Taking $\calF$ to be the set $\calF_d$ defined above, with $d\geq 3$,
we find that the restriction of $\napp$ to instances of degree at most
$d$ is equivalent to $\nCSP(\{\widehat F\mid F\in\calF_d\})$.  By
Theorem \ref{thm:nappalg} this problem has an FPRAS for each $d$.  In
this sense, $\napp$ generalises $\nSFO$ to a $\nCSP$.  Note that
$\widehat{\sigOdd_1}(0)=1/ {\sqrt 2}$ and
$\widehat{\sigOdd_1}(0)=-1/{\sqrt 2}$.  So we get a class of FPRASes
for $\nCSP$s using functions with mixed signs.

\subsection{Techniques}

Like Bubley and Dyer we will use Markov chains, but to bound the
mixing time we will instead apply the canonical paths technique.  More
precisely, we will use a multicommodity flow with cycle-unwinding as
used by Jerrum and Sinclair \cite{approxperm}. They proved the
following relevant result: for any polynomial $p$ we can sample
efficiently from the uniform distribution of perfect matchings, in
graphs $G$ satisfying
\begin{align} \frac{\text{number of matchings of order $\frac 1 2 |V(G)|-1$}}
{\text{number of matchings of order $\frac 1 2 |V(G)|$}} \leq
p(|V(G)|).
\label{eq:npm}
\end{align}

Recall that a \emph{matching} of a graph is a set of edges not sharing
any vertices, and a matching is \emph{perfect} if it has order
$|V(G)|/2$.  A perfect matching is a satisfying assignment to a
certain system of constraints: each edge is either IN or OUT, and
every variable enforces a perfect matchings constraint, that exactly
one of its incident edges is IN.  From this perspective a natural
question is: what \sigs{} can we use instead of perfect matchings
constraints?  We show that Jerrum and Sinclair's result generalises in
a certain sense to \emph{windable} functions, defined as follows.

\begin{definition}
For any finite set $J$ and any configuration $\vecx\in\{0,1\}^J$
define $\Matchp{\vecx}$ to be the set of partitions of $\{i\mid
x_i=1\}$ into pairs and singletons.  A function $F\sigf J$ is \defn{windable} if
there exist values $B(\vecx,\vecy,M)\geq 0$ for all
$\vecx,\vecy\in\{0,1\}^J$ and all $M\in\Matchp{\vecx\oplus\vecy}$
satisfying:
\begin{enumerate}
\item $F(\vecx)F(\vecy)=\sum_{M\in\Matchp{\vecx\oplus\vecy}} B(\vecx,\vecy,M)$
for all $\vecx,\vecy\in\{0,1\}^J$, and
\item
  $B(\vecx,\vecy,M)=B(\vecx\oplus\mathbf S,\vecy\oplus\mathbf S,M)$
  for all $\vecx,\vecy\in\{0,1\}^J$ and all $S\in M\in
  \Matchp{\vecx\oplus\vecy}$.
\end{enumerate}

Here $\vecx\oplus \mathbf S$ denotes the vector obtained by changing
$x_i$ to $1-x_i$ for the one or two elements $i$ in $S$.
\end{definition}

The next question is: what kinds of constraints guarantee a bound like
\eqref{eq:npm}?  We give one answer: \emph{strictly terraced}
functions.

\begin{definition}
A function $F\sigf J$ is \defn{strictly terraced} if
\[ F(\vecx)=0 \implies F(\vecx\oplus\eee i)=F(\vecx\oplus\eee j)\qquad\text{ for all $\vecx\in\{0,1\}^J$ and all $i,j\in J$}.\]
Here $\vecx\oplus \eee i$ denotes the vector obtained by
changing $x_i$ to $1-x_i$.
\end{definition}

We will discuss these definitions more throughout the paper.  Using
properties of these classes, we will establish Theorem
\ref{thm:nappalg}.  A feature of the techniques is that they cannot be
extended by expressibility reductions, where we just substitute a
constraint by a ``circuit'', a gadget gluing together other
constraints.  The following theorem makes this precise.

\begin{theorem}\label{thm:nappexpr}
Let $\calF$ be the class of strictly terraced windable functions. Then
\begin{itemize}
  \item $\calF$ is closed under taking \sigs{} of connected circuits
  \item $\calF$ contains $\sigEven_k$, $\sigOdd_k$, and $\sigNAE_k$
    for all $k\geq 1$
  \item for all finite subsets $\calF'\subset\calF$ there is an FPRAS for $\Holant(\calF')$
\end{itemize}
\end{theorem}

The reason to take $\calF'$ to be finite is to make sense of the
computational problem $\Holant(\calF')$. As in Theorem
\ref{thm:nappalg}, if one is careful about how the input is specified,
it is also possible to allow infinite $\calF'$ in some cases.

\subsection{Matching circuits}

In Section \ref{sec:mgadg} we will consider a natural type of gadget
for reducing Holant problems to $\nPM$, the problem of counting the
number of perfect matchings in a graph.

$\nPM$ is, famously, $\nP$-complete even when restricted to bipartite
instances \cite{Val79}.  This suggests that there is no efficient
exact algorithm, leaving the question of whether there is an
approximation algorithm.  A major result in this direction is that
there is an FPRAS for $\nPM$ restricted to bipartite graphs
\cite{JSV}.  Our study of matching circuits is an attempt to identify
which Holant problems reduce to $\nPM$ in the sense of expressibility.

 Consider a clique of order
four, where at the $i$'th vertex we attach an ``outgoing'' edge $d_i$.
For each of the sixteen possible subsets $M\subseteq\{d_1,d_2,d_3,d_4\}$ of the outgoing edges, we
can count the number $F(M)$ of ways to add internal edges to $F$ to
obtain a perfect matching. Because the clique of order four has $3$
perfect matchings, we have $F(\emptyset)=3$, while $F(\{d_1\})=0$ and
$F(\{d_1,d_2,d_3,d_4\})=1$.

We will say that a function $F\sigf J$ has a matchings circuit if
there is a similar graph fragment, with outgoing edges $J$, and such
that $F(\vecx)$ is the number of perfect matchings containing the
outgoing edges $\{i\in J\mid x_i=1\}$.  Substituting each vertex of a
$\Holant(\{F\})$ instance by the graph fragment gives a reduction from
$\Holant(\{F\})$ to $\nPM$.  Actually, in Section \ref{sec:mgadg}, following Jerrum and Sinclair we
will allow non-negative edge-weights and a ``fugacity'' at each
vertex, because these do not add any more computational power; the
important property is:

\begin{proposition}\label{prop:pmexpr}
If $\calF$ is a finite set of \sigs{} that have matchings circuits,
then $\Holant(\calF)\APred\nPM$.
\end{proposition}

Here $\APred$ denotes a type of approximation-preserving reduction
used to study the relative complexity of approximate counting problems
- see Section \ref{sec:compdef}. In particular, if
$\Holant(\calF)\APred\nPM$ and $\nPM$ has an FPRAS then
$\Holant(\calF)$ has an FPRAS.  The main result is the following
theorem.

\begin{theorem}\label{thm:arity3}
Let $F\sigf 3$. The following are equivalent:
\begin{enumerate}
\item $F$ is windable
\item For all $x_1,x_2,x_3\in\{0,1\}$ we have
\begin{align*}
&&&F(x_1,x_2,x_3)F(1-x_1,1-x_2,1-x_3)\\
&\leq &&F(x_1,x_2,1-x_3)F(1-x_1,1-x_2,x_3)\\
&+&&F(x_1,1-x_2,x_3)F(1-x_1,x_2,1-x_3)\\
&+&&F(x_1,1-x_2,1-x_3)F(1-x_1,x_2,x_3)
\end{align*}
\item $F$ has a matchings circuit
\end{enumerate}
\end{theorem}

Theorem \ref{thm:arity3} gives a class of problems that reduce to
counting perfect matchings.  For example, the Holant problem allowing
only the relation $\{(0,0,0),(1,0,0),(0,1,0),(1,0,1),(0,1,1)\}$
reduces to $\nPM$, but is not known to have an FPRAS.

\subsection{Related work}

A matroid is \emph{sbo (strongly basis orderable)} \cite{sbo} if for all
bases $A$ and $B$ there is a bijection $\pi\colon A\setminus B\to B\setminus
A$ such that for all $X\subseteq A\setminus B$ the set $(A\cup
\pi(X))\setminus X$ is a basis. Bouchet and Cunningham generalised the sbo
property as \emph{linkability} for the class of even delta-matroids, and
showed that this class is closed under an analogue of circuits
\cite{linkable}. These conditions are just windability over the
two-element Boolean semiring $(\mathbb{B}=\{0,1\},\max,\min)$, for the set of
bases when considered as a function $\{0,1\}^J\to\mathbb{B}$,
by taking the characteristic vector of
characteristic vectors of bases.  Gambin used the sbo property to
approximately count the number of bases in certain matroids
\cite{Gambin}.

Valiant \cite{accident} introduced \emph{matchgates} and
\emph{matchcircuits}, which are similar to matchings circuits but give
efficient exact algorithms. Matchcircuits can be understood as planar
graphs with edge-weights, with no restriction to non-negative numbers. Cai
and Choudhary characterised the expressibility of matchgates
\cite{choudhary}.  The name ``matchings circuits'' used in this paper
is meant to suggest a version of matchgates.

The focus on (the negative side of) expressibility for approximate
counting problems appears in \cite{lsm}, where
logsupermodular functions are shown not to express non-logsupermodular
functions in the context of $\nCSP$s.

Yamakami \cite{Yamakami} and the current author \cite{cspvw} have
given partial classifications for classes of Holant problems.  The
bulk of these results deal with intractability: reductions from a
named problem such as $\nSAT$ to a given Holant problem.  The
focus of the current paper is on tractability: either in the absolute
sense of an FPRAS, or by reductions to $\nPM$.

A related $\nCSP$ with mixed signs appears in the context of the Tutte
polynomial.  By the proof of \cite[Lemma 7]{Tutte}, the following
problems are equivalent in the sense of approximate counting, for any
fixed $y<-1$.
\begin{itemize}
\item $\nPM$
\item $\nCSP(\{B_y\})$
where $B_y\sigf 2$ is defined by
$B_y(0,0)=B_y(1,1)=y$ and
$B_y(0,1)=B_y(1,0)=1$
\item evaluating the Tutte polynomial at the point $(x,y)$ where $(x-1)(y-1)=2$
\end{itemize}

\subsection{Outline}

In Section \ref{sec:evenwind}, we adapt the conductance argument of
Jerrum and Sinclair to ``even-windable'' functions, which are a
slightly simpler version of windable functions. 
We study windable functions
in Section \ref{sec:windable}.  We study strictly terraced functions
in Section \ref{sec:strterr}.  In Section \ref{sec:mainproofs} we
establish Theorem \ref{thm:nappalg} and Theorem \ref{thm:nappexpr}.
Finally, in Section \ref{sec:mgadg} we discuss matchings circuits and
establish Proposition \ref{prop:pmexpr} and Theorem~\ref{thm:arity3}.

\section{Preliminaries}\label{sec:prelim}

A \defn{configuration} of a finite ``indexing'' set $J$ is a
function $\vecx\in\{0,1\}^J$.  A \defn{\sig{}} is a function $F\sigf
J$.  A \defn{copy} of $F$ is a function $G\sigf I$ of the form
$G(\vecx)=F(\vecx\circ \pi)$ for some bijection $\pi\colon J\to I$.  We will use
bold face to distinguish between sets $S\subset J$ and the
characteristic vector $\mathbf{S}$.  Similarly the bold version of a
relation $R\subseteq\{0,1\}^J$ is the corresponding zero-one-valued
\sig{}.

We will not distinguish between $\{0,1\}^{\{1,\dots,k\}}$ and
$\{0,1\}^k$, or between $\{0,1\}^1$ and $\{0,1\}$.  Also, we will
sometimes allow indexing sets to be partially enumerated in a certain
way.  This is for notational power: the enumerated indices are easy
to refer to explicitly, while the unenumerated indices are easy to
fix. For all positive integers $k$ and all finite sets $J$, when
$k+J$ is used as an indexing set it means the disjoint union of
$\{1,\dots,k\}$ and $J$.  Elements of $\{0,1\}^{k+J}$ will be denoted
by $(x_1,\dots,x_k;\vecy)$ where $x_1,\dots,x_k\in\{0,1\}$ and
$\vecy\in\{0,1\}^J$.

For all sets $I\subseteq J$, all configurations $\vecp$ of $I$ and
all configurations $\vecx$ of $J\setminus I$, let $(\vecx,\vecp)$
denote the unique common extension of $\vecx$ and $\vecp$ to a
configuration of $J$.  The \defn{pinning} of a \sig{} $F\sigf J$ by
$\vecp\in\{0,1\}^I$ ($I\subseteq J$) is the \sig{} $F'\sigf {J\setminus I}$
defined by $F'(\vecx)=F(\vecx,\vecp)$.

The \defn{distance} $|\{i\in J\mid x_i\neq y_i\}|$ between two
configurations $\vecx,\vecy\in\{0,1\}^J$ will be denoted
$\dist{\vecx}{\vecy}$.  We say $F$ is \defn{even}\footnote{This may be
  confusing terminology - an even function can be non-zero on vectors
  of \emph{odd} Hamming weight. But it is a common definition for
  delta-matroids.}  if $\dist{\vecx}{\vecy}$ is even for all
$\vecx,\vecy$ with $F(\vecx),F(\vecy)>0$.  Define
$\vecx\oplus\vecy\in\{0,1\}^J$ by $(\vecx\oplus\vecy)_i\equiv
x_i+y_i\pmod 2$.  For all $\vecx\in\{0,1\}^J$ define
$\overline{\vecx}\in\{0,1\}^J$ by $\overline{x}_i=1-x_i$, and for all
$F\sigf J$ define $F\overline F\sigf J$ by $F\overline
F(\vecx)=F(\vecx)F(\overline{\vecx})$.  For all $F\sigf J$ and
$\vecy\in\{0,1\}^J$ define the \defn{flip of $F$ by $\vecy$} to be the
\sig{} $F'\sigf J$ defined by $F'(\vecx\oplus\vecy)=F(\vecx)$ for all
$\vecx\in\{0,1\}^J$.  For all $i\in J$ define $\eee i\in\{0,1\}^J$
(where $J$ is implicit) to be the characteristic vector of $\{i\}$.

For all finite sets $J$ define
\begin{align*}
\relEven_J&=\{\vecx\in\{0,1\}^J\mid \sum_{i\in J}x_i\text{ is even}\}\\
\relOdd_J&=\{\vecx\in\{0,1\}^J\mid \sum_{i\in J}x_i\text{ is odd}\}\\
\relNAE_J&=\{\vecx\in\{0,1\}^J\mid 1\leq \sum_{i\in J}x_i\leq |J|-1\}\\
\relEvenNAE_J&=\relEven_J\cap \relNAE_J
\end{align*}
$\relEven_J$ and $\relOdd_J$ are \defn{parity relations}.  The last
relation $\relEvenNAE_J$ is only used for calculations (and only with
$|J|$ even).

\subsection{Circuits}

In this paper, circuits are a type of graph equipped with \sigs{} at
each vertex, and allowing external edges. A little care is needed to
allow self-loops and asymmetric \sigs{}.

A \defn{graph fragment $G$} is specified by:
\begin{itemize}
\item a set $J^G$ whose elements are called incidences
\item a set $V^G$ of vertices, and sets $J_v^G$, $v\in V^G$, that partition $J^G$
\item a set $A^G\subseteq J^G$ whose elements are called external edges
\item a partition $E^G$ of $J^G\setminus A^G$ into pairs called internal edges
\end{itemize}

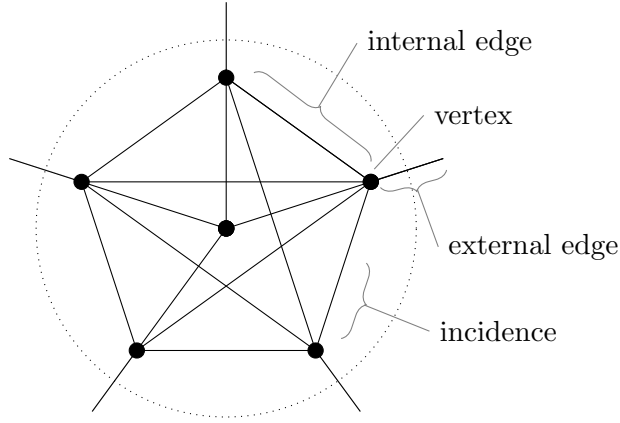
\begin{figure}
  \centering
\begin{tikzpicture}
  \draw[dotted] (0:0) circle (2.5cm);
  \foreach \i in {0,1,2,3,4,5} \draw (18+72*\i:2) -- (18+72*1+72*\i:2);
  \foreach \i in {2,3,4,5} \draw (18+72*\i:2) -- (18+72*2+72*\i:2);
  \foreach \i in {0,1,3,4,5} \fill (18+72*\i:2);
  \foreach \i in {0,1,2,3} \draw (18+72*\i:2) -- (0,0) circle (0.1cm);
  \foreach \i in {0,1,2,3,4,5} \filldraw (18+72*\i:2) circle (0.1cm) -- (18+72*\i:3);
  \fill (0,0) circle (0.1cm);

  \draw [gray,decorate,decoration={brace,amplitude=5pt,mirror}]
     (25:2.1) -- (80:2.1) node (intedge) [midway] {};
  \draw [gray,decorate,decoration={brace,amplitude=5pt,mirror}]
     (28-72:2.1) -- (58-72:1.9) node (halfedge) [midway] {};
  \draw [gray,decorate,decoration={brace,amplitude=5pt,mirror}]
     (15:2.1) -- (15:3) node (extedge) [midway] {};
  
  \draw[gray]  (intedge) -- (55:3) node[anchor=west,black] {internal edge};
  \draw[gray] (halfedge) -- (45-72:3) node[anchor=west,black] {incidence};
  \draw[gray] (extedge) -- (-5:2.8) node[anchor=west,black] {external edge};

  \draw[gray] (30:3) node[anchor=west,black] {vertex} -- (22:2.1);
\end{tikzpicture}
\caption{Terminology for graph fragments.}\label{fig:fragment}
\end{figure}

See Figure \ref{fig:fragment}.

A \defn{circuit} $\phi$ is graph fragment equipped with a
\defn{constraint} $F^\phi_v\sigf {J_v^\phi}$ for each vertex $v$.
We can also use a relation $R\subseteq\{0,1\}^{J_v^\phi}$ as a constraint
by taking $F^\phi_v=\mathbf R$.
We
will drop the superscript $\phi$ where there is only one graph or circuit in
context.

$G$ is \defn{closed} if it has no external edges.
Standard graph-theoretic terminology extends to graph fragments. In particular
we will refer to connected graph fragments.
An \emph{edge} is either an internal edge or an external edge.
A vertex $v$ and an internal
edge $e$ are \emph{incident} if $J_v$ intersects $e$.  If an internal edge $e$
is uniquely identified by the vertices $u,v$ it is incident to, we will
denote $e$ by $uv$.

Given a circuit $\phi$, for any configuration $\vecx$ of $J$,
\begin{itemize}
\item $\vecx$ is an \defn{assignment} (with respect to $E$) if $x_i=x_j$ for all $\{i,j\}\in E$.
\item $\vecx|_{J_v}$ denotes the restriction of $\vecx$ to $J_v$.
\item The \defn{weight} of $\vecx$ is $\wt_\phi(\vecx)=\prod_{v\in V}
  F_v(\vecx|_{J_v})$.
\end{itemize}

The \defn{\sig{} of $\phi$} is the function $\sem{\phi}\sigf A$
defined by
$$\sem{\phi}(\vecx)=\sum_{\vecx'}\wt_\phi(\vecx')\qquad\text{
  ($\vecx\in\{0,1\}^A$)}$$ where the sum is over extensions of
$\vecx$ to assignments $\vecx'\sigf J$ with respect to $E$.  If a \sig{} $F$ is equal to
$\sem{\phi}$, we will say that $F$ \defn{has} the circuit $\phi$.

Another way to think of a circuit is as a ``read-twice
pps-formula'', a special case of the pps-formulas of \cite{lsm}.  For
example, consider an equation
\[ F(x)=\sum_{y=0}^1 \sum_{z=0}^1 G(x,y)G(y,z)H(z)\qquad (x\in\{0,1\}). \]
Note how on the right-hand-side, each bound (summed) variable appears
exactly twice, and each free (unsummed) variable appears exactly
once. Any equation of this form defines a circuit in a natural
way: incidences correspond to the variable occurrences $x,y,y,z,z$;
vertices correspond to terms $G(x,y),G(y,z),H(z)$; the sets $J_v$ are
scopes for each term; external edges correspond to free variables; and
internal edges correspond to summed variables.

For any partition $E$ of a finite set $J$ into pairs,
for all non-negative integers $k$, a \defn{$k$-assignment} with respect to $E$ is a configuration $\vecx$ of $J$ such that
$x_i=x_j$ for all but exactly $k$ pairs $\{i,j\}\in E$.  So an
assignment is a $0$-assignment.  For all closed circuits $\phi$ and
all integers $k\geq 0$ define
$$Z_k(\phi)=\sum_{\text{$k$-assignments $\vecx$}}\wt_\phi(\vecx).$$
So $Z_0(\phi)$ is just $\sem{\phi}$ (evaluated on the empty
configuration).

\subsection{Computational definitions}\label{sec:compdef}

A \emph{counting problem} is a function $f$ taking instances (encoded
as strings over a finite alphabet $\Sigma$) to non-negative reals.  A
\emph{randomised approximation scheme\/} for $f$ is a randomised
algorithm that takes an instance $x$ and error parameter $\epsilon>0$
and returns an approximation $Z$ to $f(x)$ satisfying
\begin{equation}
\Pr \big[e^{-\epsilon}f(x)\leq Z \leq e^\epsilon f(x)\big]\geq 3/4.
\end{equation}

 A \emph{fully polynomial randomised approximation scheme (FPRAS)} for
 $f$ is a randomised approximation scheme that runs in polynomial time
 in $|x|$ and $\epsilon^{-1}$.  (To be concrete, we can require the
 error parameter to be specified by a binary integer $\epsilon^{-1}$.)

Let $f$ and $g$ be counting problems.  A randomised oracle algorithm
$\mathcal{A}$ meeting the following conditions is an
\emph{approximation-preserving reduction} from $f$ to $g$, and if such
a reduction exists we write $f\APred g$.

$\mathcal{A}$ takes inputs
$(w,\epsilon)$ where $w$ is in the domain of $f$, and
$\epsilon>0$. The run-time of $\mathcal{A}$ is polynomial in $|w|$ and
$\epsilon^{-1}$ and the bit-size of the values returned by the oracle
(this avoids requiring that the oracle gives concise responses).  The
oracle calls made by $\mathcal{A}$ are of the form $(v,\delta)$, where
$v$ is an instance of $g$ and $\delta>0$ is an error parameter, such
that $|v|+\delta^{-1}$ is bounded by a polynomial in $|w|$ and
$\epsilon^{-1}$ (depending only on $\mathcal{A}$).  If the oracle's
outputs meet the specification of a randomised approximation scheme
for $g$, then $\mathcal{A}$ is a randomised approximation scheme for
$f$.

The above definitions are based on \cite{DGGJ}; the main difference is
that we allow non-integer-valued problems.

For any finite set $\calF$ of \sigs{} define the following
counting problem.
\prob{$\Holant(\calF)$}
{A closed circuit $\phi$ using copies of \sigs{} in $\calF$}
{$\sem{\phi}$}

Since $\calF$ is finite, it is not particularly important how the
functions $F^\phi_v$ are specified. For concreteness: $F^\phi_v$
should be specified by an index $i$ into a fixed enumeration
$\calF=\{F_1,\dots,F_{|\calF|}\}$, along with a bijection from
$J^\phi_v$ to the indexing set $I_i$ of $F_i\sigf {I_i}$.

By substituting circuits, if $F$ has a circuit
using copies of \sigs{} from a finite set $\calF$, then
$\Holant(\calF\cup\{F\})\APred\Holant(\calF)$.  This
justifies the focus on expressibility in this paper.

\section{Even-windable functions}\label{sec:evenwind}

\subsection{Idea}

Windability is an abstraction of a property of the distribution of
perfect matchings in a graph with external edges.  We will illustrate
the idea briefly by the arity $4$ case, where windability is already
used implicitly in \cite{approxperm}. But higher-arity conditions are
important for showing that windability is preserved by circuits.

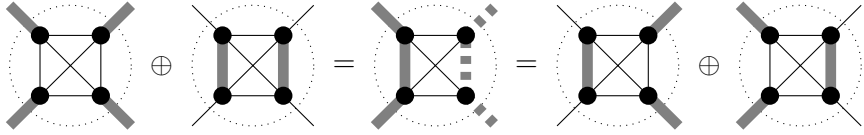
\begin{figure}
\begin{tikzpicture}[scale=0.4,
  ]
  \begin{scope}
  \draw[dotted] (2,2) circle (2);
  \draw (1,1)--(1,3)--(3,1)--(3,3)--(1,1)--(3,1);
  \draw (1,3)--(3,3);
  \draw[spun] (1,1) -- (0,0);
  \draw[spun] (1,3) -- (0,4);
  \draw[spun] (3,1) -- (4,0);
  \draw[spun] (3,3) -- (4,4);
  \foreach \x/\y in {1/1,3/3,1/3,3/1} \drawvertb{\x}{\y};
  \end{scope}
  \draw node at (5,2) {$\oplus$};
  \draw node at (11,2) {$=$};
  \draw node at (17,2) {$=$};
  \draw node at (23,2) {$\oplus$};
  \begin{scope}[xshift=6cm]
  \draw[dotted] (2,2) circle (2);
  \draw[spun] (1,1)--(1,3);
  \draw[spun] (3,1)--(3,3);
  \draw (1,1)--(3,1)--(1,3)--(3,3)--(1,1);
  \draw (1,1) -- (0,0);
  \draw (1,3) -- (0,4);
  \draw (3,1) -- (4,0);
  \draw (3,3) -- (4,4);
  \foreach \x/\y in {1/1,3/3,1/3,3/1} \drawvertb{\x}{\y};
  \end{scope}
  \begin{scope}[xshift=12cm]
  \draw[dotted] (2,2) circle (2);
  \draw[spun] (1,1)--(1,3);
  \draw[dashed, spun] (3,1)--(3,3);
  \draw (1,1)--(3,1)--(1,3)--(3,3)--(1,1);
  \draw[spun] (1,1) -- (0,0);
  \draw[spun] (1,3) -- (0,4);
  \draw[dashed, spun] (3,1) -- (4,0);
  \draw[dashed, spun] (3,3) -- (4,4);
  \foreach \x/\y in {1/1,3/3,1/3,3/1} \drawvertb{\x}{\y};
  \end{scope}

  \begin{scope}[xshift=18cm]
  \draw[dotted] (2,2) circle (2);
  \draw[spun] (1,1)--(1,3);
  \draw (3,1)--(3,3);
  \draw (1,1)--(3,1)--(1,3)--(3,3)--(1,1);
  \draw (1,1) -- (0,0);
  \draw (1,3) -- (0,4);
  \draw[spun] (3,1) -- (4,0);
  \draw[spun] (3,3) -- (4,4);
  \foreach \x/\y in {1/1,3/3,1/3,3/1} \drawvertb{\x}{\y};
  \end{scope}

  \begin{scope}[xshift=24cm]
  \draw[dotted] (2,2) circle (2);
  \draw (1,1)--(1,3);
  \draw[spun]  (3,1)--(3,3);
  \draw (1,1)--(3,1)--(1,3)--(3,3)--(1,1);
  \draw[spun] (1,1) -- (0,0);
  \draw[spun] (1,3) -- (0,4);
  \draw (3,1) -- (4,0);
  \draw (3,3) -- (4,4);
  \foreach \x/\y in {1/1,3/3,1/3,3/1} \drawvertb{\x}{\y};
  \end{scope}
\end{tikzpicture}

\caption{An example of constructing perfect matchings by symmetric
  differences. From left to right, $M$, $M'$, $M\triangle M'$ (with $P$ drawn
  in thick solid grey),
  $M\triangle P$, and $M'\triangle P$.}\label{fig:open}
\end{figure}

Consider a graph $G$ with four external edges $e_1,e_2,e_3,e_4$.  For
all $x_1,x_2,x_3,x_4$, let $F(x_1,x_2,x_3,x_4)$ be the number of
perfect matchings in $G$ that include the outgoing edges $\{e_i\mid
x_i=1\}$.  So $F(0,0,0,0)F(1,1,1,1)$ is the number of pairs of perfect
matchings $(M,M')$ such that $M$ includes all the external edges and
$M'$ includes none.  But for any such pair $(M,M')$, the symmetric
difference $M\triangle M'$ consists of cycles and paths, and the path
starting at $e_1$ ends at either $e_2$, $e_3$, or $e_4$, depending on
the choice of $(M,M')$.  Thus $F(0,0,0,0)F(1,1,1,1)$ splits into three
terms.  Denote these by $B((0,0,0,0),(1,1,1,1),M)$ where $M$ is a
partition of $\{1,2,3,4\}$ into pairs: either $\{\{1,2\},\{3,4\}\}$ or
$\{\{1,3\},\{2,4\}\}$ or $\{\{1,4\},\{2,3\}\}$.  We can similarly define
  $B((1,1,0,0),(0,0,1,1),M)$ for example.

When $M\triangle M'$ contains a path $P$ from $e_1$ to $e_2$, the sets
$M\triangle P$ and $M'\triangle P$ are also perfect matchings - see
Figure \ref{fig:open}. The only external edges in $M\triangle P$ are
$e_3$ and $e_4$, while the only external edges in $M'\triangle P$ are
$e_1$ and $e_2$.  Thus $B((0,0,0,0),(1,1,1,1),\{\{1,2\},\{3,4\}\})$
equals $B((1,1,0,0),(0,0,1,1),\{\{1,2\},\{3,4\}\}$.

In this section, for simplicity, we will consider only even functions.

\subsection{Definition}

For any configuration $\vecx\in\{0,1\}^J$ define $\Match{\vecx}$ to be
the set of partitions of $\{i\in J\mid x_i=1\}$ into pairs. 
In particular, if $\sum_{i\in J}x_i$ is odd then $\Match{\vecx}=\emptyset$.

A function $F\sigf J$ is \defn{even-windable} (with witness $B$) if
there exist values $B(\vecx,\vecy,M)\geq 0$ for all
$\vecx,\vecy\in\{0,1\}^J$ and all $M\in\Match{\vecx\oplus\vecy}$,
i.e. all partitions $M$ of the set $\{i\in J\mid x_i\neq y_i\}$ into
pairs, satisfying:
\begin{enumerate}
\item[EW1.] $F(\vecx)F(\vecy)=\sum_{M\in\Match{\vecx\oplus\vecy}} B(\vecx,\vecy,M)$
for all $\vecx,\vecy\in\{0,1\}^J$, and
\item[EW2.]
  $B(\vecx,\vecy,M)=B(\vecx\oplus\mathbf S,\vecy\oplus\mathbf S,M)$
  for all $\vecx,\vecy\in\{0,1\}^J$ and all $S\in M\in \Match{\vecx\oplus\vecy}$.
\end{enumerate}
Note that in the second condition, $S$ is a pair $\{i,j\}$ in $M$: we
are swapping the values of $x_i$ and $y_i$, and swapping the values of
$x_j$ and $y_j$.
By swapping a sequence of pairs, EW2 is equivalent to

\begin{enumerate}
\item[EW2'.]
$B(\vecx,\vecy,M)=B(\vecx\oplus\mathbf S_1\oplus\dots\oplus\mathbf S_k,\vecy\oplus\mathbf S_1\oplus\dots\oplus\mathbf S_k,M)$ for
all $\vecx,\vecy\in\{0,1\}^J$ and all $S_1,\dots,S_k\in M\in \Match{\vecx\oplus\vecy}$.
\end{enumerate}

\subsection{2-decompositions}\label{sec:evenwind_examples}

Using pinnings, the even-windability conditions can be stated in a form
that is sometimes easier to check. A function $H\sigf J$ has a
\defn{2-decomposition} if there are values $D(\vecx,M)\geq 0$, where
$\vecx$ ranges over $\{0,1\}^J$ and $M$ ranges over partitions of $J$
into pairs, such that:
\begin{enumerate}
\item $H(\vecx)=\sum_M D(\vecx,M)$ for all $\vecx$, where the sum is over partitions of $J$ into pairs, and
\item $D(\vecx,M)=D(\vecx\oplus\mathbf S,M)$ for all $\vecx,M$ and all $S\in M$.
\end{enumerate}
In particular if $|J|$ is odd then the first condition forces $H$ to be identically zero.

A function $F$ is even-windable if and only if for all pinnings $G$ of
$F$ the function $G\overline{G}$ has a 2-decomposition.  For the
forwards direction, given a witness $B$ that $F$ is even-windable, for
each $I\subseteq J$ and each $\vecp\in\{0,1\}^I$ define
$D_{\vecp}(\vecx,M)=B((\vecx,\vecp),(\overline{\vecx},\vecp),M)$ for
all $\vecx\in\{0,1\}^{J\setminus I}$ to obtain a 2-decomposition
$D_{\vecp}$ of the pinning of $F$ by $\vecp$.  For the backwards
direction, for each $I\subseteq J$ and each $\vecp\in\{0,1\}^I$, pick
a 2-decomposition $D_{\vecp}$ of the pinning of $F$ by $\vecp$.  For
all $\vecx,\vecy\in\{0,1\}^J$, define
$B(\vecx,\vecy,M)=D_\vecp(\vecx',M)$ where $\vecp$ is the restriction
of $\vecx$ to $\{i\in J\mid x_i=y_i\}$ and $\vecx'$ is the restriction
of $\vecx$ to $\{i\in J\mid x_i\neq y_i\}$. Then $B$ witnesses that $F$ is
even-windable.

\begin{lemma}\label{lem:evenwind_small}
Let $F\sigf J$ with $|J|\leq 3$. If $F$ is even then $F$ is
even-windable.
\end{lemma}
\begin{proof}
Let $G\sigf I$ be a pinning of $F$.

If $I=\emptyset$ define
$D(\vecx,\emptyset)=G\overline G(\vecx)$ where $\vecx\in\{0,1\}^\emptyset$ is
the empty configuration.  Then $G\overline G(\vecx)=\sum_M D(\vecx,M)$
where $M$ ranges over the set $\{\emptyset\}$ of partitions of $I$
into pairs, so $D$ is a 2-decomposition of $G\overline G$.

If $|I|=2$, let $i,j$ be the elements of $I$ and define
$D(\vecx,\{\{i,j\}\})=G\overline G(\vecx)$.  For all
$\vecx\in\{0,1\}^I$ we have $G\overline G(\vecx)=\sum_M D(\vecx,M)$ where $M$
ranges over the set $\{\{\{i,j\}\}\}$ of partitions of $I$ into pairs,
so $D$ is a 2-decomposition of $G\overline G$.

If $|I|$ is $1$ or $3$ then $G(\vecx)$ and $G(\overline{\vecx})$
cannot be simultaneously be non-zero because $G$ is a pinning of the
even function $F$, and $\sum_{i\in I}x_i\equiv |I|+\sum_{i\in
  I}(1-x_i)\pmod 2$. Thus $G\overline G$ is identically zero.  There
are also no partitions of $I$ into pairs, so the empty function is a
2-decomposition of $G\overline G$.
\end{proof}

\begin{lemma}\label{lem:evenoddeven}
$\sigEven_J$ and $\sigOdd_J$ have a 2-decomposition
whenever $|J|$ is even.  $\sigEven_J$ and $\sigOdd_J$ are
even-windable for any $J$.
\end{lemma}
\begin{proof}
First consider $\sigEven_J$.  Fix a partition $N$ of $J$ into pairs.  Define
$$D(\vecx,M)=\begin{cases}1&\text{ if $M=N$ and $\sum_{i\in J}x_i$ is even}\\
0&\text{ otherwise.}
\end{cases}$$
Then for all $\vecx\in\{0,1\}^J$ we have
$\sigEven_J(\vecx)=\sum_M D(\vecx,M)$ (where the sum ranges over
partitions $M$ of $J$ into pairs). Similarly for $\sigOdd_J$, define
$$D(\vecx,M)=\begin{cases}1&\text{ if $M=N$ and $\sum_{i\in J}x_i$ is odd}\\
0&\text{ otherwise.}
\end{cases}$$
Then for all $\vecx\in\{0,1\}^J$ we have
$\sigOdd_J(\vecx)=\sum_M D(\vecx,M)$.

Now consider a pinning $G\sigf K$ of $\sigEven_J$ or $\sigOdd_J$.
If $|K|$ is odd then $G\overline G$ is identically zero, by the same
argument used in Lemma \ref{lem:evenwind_small}.  Otherwise,
$G=G\overline G$ is either $\sigEven_K$ or
$\sigOdd_K$, which we showed have 2-decompositions.
\end{proof}

The following argument gives a more difficult example of a
2-decomposition. It will be used later (in the proof of Lemma
\ref{lem:flipdecomp}) to show that $\sigNAE_J$ is windable.

\begin{lemma}\label{lem:evenwind_paritynae}
Let $J$ be an finite set with $|J|$ even.  Then $\sigEvenNAE_J$ has
a 2-decomposition.
\end{lemma}
\begin{proof}
For each subset $I\subseteq J$ of even order fix a partition $M_I$ of
$I$ into pairs. Set
$$D(\vecx,M)=2^{-k+2}\left|\{I\subseteq J\mid \text{$|I|$ is even, $\sum_{i\in I}x_i$ and $\sum_{i\in J\setminus I}x_i$ are odd, and $M=M_I\cup M_{J\setminus I}$}\}\right|.$$

$S\in M_I\cup M_{J\setminus I}$ implies $S\subseteq I$ or $S\subseteq
J\setminus I$.  The conditions that $\sum_{i\in I}x_i$ and $\sum_{i\in
  J\setminus I}x_i$ are odd are therefore not affected by changing
$\vecx$ to $\vecx\oplus\mathbf S$.  Thus $D(\vecx\oplus\mathbf
S,M)=D(\vecx,M)$ for all $S\in M$.

For any $\vecx$, if $\sigEvenNAE_J(\vecx)=0$ then $D(\vecx,M)=0$. If
$\sigEvenNAE_J(\vecx)=1$, pick $i,j$ with $x_i=0$ and $x_j=1$. For
each of the $2^{k-2}$ subsets $I'\subseteq J\setminus\{i,j\}$ there is
a unique set $I''\subseteq\{i,j\}$ such that the order of $I=I'\cup
I''$ is even and such that $\sum_{i\in I}x_i$ and $\sum_{i\in
  J\setminus I}x_i$ are odd. There are thus $2^{k-2}$ such subsets
$I$ for each fixed $\vecx$, which gives $\sum_M D(\vecx,M)=1$. So $D$
is a 2-decomposition of $\sigEvenNAE_J$.
\end{proof}

\subsection{Expressibility}\label{sec:ewexpr}

We will show that the \sig{} of any circuit using even-windable
functions is even-windable.  We will use a certain graph
associated with a choice of matching of incidences.

Let $M$ and $E$ each be a set of disjoint pairs of some set.  Define
the \defn{link graph} $L_E(M)$ to be the multigraph on the vertex set
$\bigcup_{S\in M} S$ with edge set the disjoint union of $M$ and
$\{\{i,j\}\in E \mid i,j\in \bigcup_{S\in M}S\}$ (so edges in $M\cap
E$ give pairs of parallel edges in $L_E(M)$).

Note that for each vertex $i$ of $L_E(M)$, the degree of $i$ is two if
$\{i,j\}\in E$ for some $j\in\bigcup_{S\in M}S$, and otherwise $i$ has
degree one.  So $L_E(M)$ consists of paths and cycles.

\begin{figure}
\begin{tikzpicture}

\foreach \x/\y in {0/0,0/2,2/0,2/2} {
\draw (\x,\y) circle (0.6cm);
}

\draw
(0.2,-0.2) -- (1.8,-0.2)
(2.2,0.2) -- (2.2,1.8)
(0.2,2.2) -- (1.8,2.2)
(-0.2,0.2) -- (-0.2,1.8)
(0.2,0.2) -- (1.8,1.8);
\draw (-0.2,-0.2) -- ++(-1,-1);
\draw (-0.2,2.2) -- ++(-1,1);
\draw (2.2,-0.2) -- ++(1,-1);
\draw (2.2,2.2) -- ++(1,1);

\draw[line width=0.1cm]
(-0.2,-0.2) -- (0.2,-0.2)
(-0.2,0.2) -- (0.2,0.2)
(1.8,-0.2) -- (2.2,-0.2)
(-0.2,1.8) -- (0.2,2.2)
(1.8,1.8) -- (1.8,2.2);

\begin{scope}[xshift=0cm,yshift=0cm]
\drawincg {-0.2}{-0.2} ;
\drawincg {0.2}{-0.2} ;
\drawincg {0.2}{0.2} ;
\drawincg {-0.2}{0.2} ;
\end{scope}

\begin{scope}[xshift=0cm,yshift=2cm]
\drawincg {0.2}{0.2} ;
\drawincg {-0.2}{-0.2} ;
\drawinc {-0.2}{0.2} ;
\end{scope}

\begin{scope}[xshift=2cm,yshift=2cm]
\drawinc {0.2}{0.2} ;
\drawincg {-0.2}{-0.2} ;
\drawincg {-0.2}{0.2} ;
\drawinc {0.2}{-0.2} ;
\end{scope}

\begin{scope}[xshift=2cm,yshift=0cm]
\drawinc {0.2}{0.2} ;
\drawincg {-0.2}{-0.2} ;
\drawincg {0.2}{-0.2} ;
\end{scope}

\draw[double, thick, ->] (4,1) -- (6,1);

\begin{scope}[xshift=8cm]

\draw[line width=0.1cm]
(-0.2,-0.2) -- (0.2,-0.2)
(-0.2,0.2) -- (0.2,0.2)
(1.8,-0.2) -- (2.2,-0.2)
(-0.2,1.8) -- (0.2,2.2)
(1.8,1.8) -- (1.8,2.2);

\draw
(0.2,2.2) -- (1.8,2.2)
(-0.2,0.2) -- (-0.2,1.8)
(0.2,0.2) -- (1.8,1.8)
(0.2,-0.2) -- (1.8,-0.2);

\begin{scope}[xshift=0cm,yshift=0cm]
\drawincg {-0.2}{-0.2} ;
\drawincg {0.2}{-0.2} ;
\drawincg {0.2}{0.2} ;
\drawincg {-0.2}{0.2} ;
\end{scope}

\begin{scope}[xshift=0cm,yshift=2cm]
\drawincg {0.2}{0.2} ;
\drawincg {-0.2}{-0.2} ;
\end{scope}

\begin{scope}[xshift=2cm,yshift=2cm]
\drawincg {-0.2}{-0.2} ;
\drawincg {-0.2}{0.2} ;
\end{scope}

\begin{scope}[xshift=2cm,yshift=0cm]
\drawincg {-0.2}{-0.2} ;
\drawincg {0.2}{-0.2} ;
\end{scope}

\end{scope}

\end{tikzpicture}
\caption{A circuit $\phi$ and a link graph $L_E(M)$ for some
  $M\in\Match{\vecx}$ where $\vecx$ is an assignment of
  $E^\phi$.
(In particular, the $M$ drawn is a union of partitions $M_v\in\Match{\vecx|_{J_v}}$.)
Circles represent vertices of the circuit.  Squares are
  incidences $i\in J^\phi$ of the circuit, and are filled black where
  $x_i=1$.  Elements of $M$ are drawn as thick black lines. Elements
  $\{i,j\}\in E$ are drawn as thin lines.\label{fig:lg1}}
\end{figure}
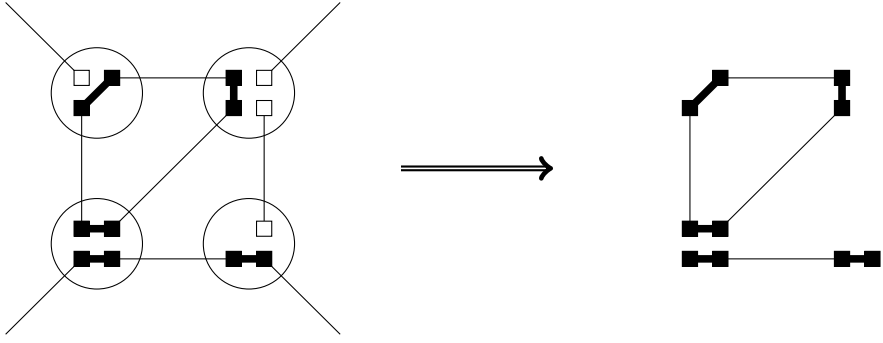

We will use this graph later for the analysis of the near-assignments
chain.  For now, consider an assignment $\vecx$ of some circuit with
internal edges $E$ and external edges $A$, and let
$M\in\Match{\vecx}$.  For any $i$ not in $A$ with $x_i=1$, the unique
$j$ with $\{i,j\}\in E$ satisfies $x_j=1$. This means that
$i\in\bigcup_{S\in M}S$ has degree $1$ in $L_E(M)$ if and only if
$i\in A$.  So every path component of $L_E(M)$ ends in $\{i\in A\mid
x_i=1\}$, and every such $i$ is at the end of a path. See Figure
\ref{fig:lg1}.

\begin{lemma}\label{lem:evenwind_expr}
Let $\phi$ be a circuit using only \sigs{} that are even-windable.
The \sig{} of $\phi$ is even-windable.
\end{lemma}
\begin{proof}
Recall that $V,J,J_v,A,E,F_v$ denote vertices, incidences, vertices'
incidences, external edges, internal edges, and vertices' \sigs{}.
For each $v\in V$ pick a function $B_v$ witnessing that $F_v$ is
even-windable.

Consider a set $M'$ of disjoint pairs of $J$. We will say
that $M'$ \emph{induces} the set of pairs $\{i,j\}\subseteq A$ such
that there is a path from $i$ to $j$ in $L_E(M')$.

For all $\vecx,\vecy\in\{0,1\}^A$ and all
$M\in\Match{\vecx\oplus\vecy}$ define
\newcommand\sumind{\sum_{\text{$\{M_v\}$ inducing $M$}} }
$$B(\vecx,\vecy,M)=\sum_{\vecx',\vecy'} \;\; \sumind \;\; \prod_{v\in V} B_v(\vecx'|_{J_v},\vecy'|_{J_v},M_v)$$
where:
\begin{itemize}
\item $\sum_{\vecx',\vecy'}$ denotes the sum over assignments
  $\vecx'$ and $\vecy'$ extending $\vecx$ and $\vecy$ respectively.
\item $\sumind$ denotes the sum over all choices of
  $M_v\in\Match{(\vecx'\oplus\vecy')|_{J_v}}$ for each
  $v\in V$, such that $\bigcup_{v\in V} M_v$ induces $M$.
\end{itemize}

For all $\vecx,\vecy\in\{0,1\}^A$ we have
\begin{align*}
\sum_{M\in\Match{\vecx\oplus\vecy}}B(\vecx,\vecy,M)
&=\sum_{M\in\Match{\vecx\oplus\vecy}} \;\; \sum_{\vecx',\vecy'} \;\; \sumind \;\; \prod_{v\in V} B_v(\vecx'|_{J_v},\vecy'|_{J_v},M_v)\\
&=\sum_{\vecx',\vecy'} \sum_{\{M_v\}} \prod_{v\in V} B_v(\vecx'|_{J_v},\vecy'|_{J_v},M_v)\\
&=\sum_{\vecx',\vecy'} \prod_{v\in V} F_v(\vecx'|_{J_v})F_v(\vecy'|_{J_v})\\
&=\sem{\phi}(\vecx) \sem{\phi}(\vecy).
\end{align*}
Here $\sum_{\{M_v\}}$ denotes the sum over \emph{all} choices of
$M_v\in\Match{(\vecx'\oplus\vecy')|_{J_v}}$ for each $v\in V$:
the sum over $M$ eliminates the condition that $\bigcup_{v\in V} M_v$
induces $M$.

Now fix $\vecx,\vecy\in\{0,1\}^A$ and $S=\{i,j\}\in
M\in\Match{\vecx\oplus\vecy}$.  For any choice of $\{M_v\}$ inducing
$M$, there is a unique path component $P_{\{M_v\}}$ (also depending on
$\vecx,\vecy,S$) from $i$ to $j$ in $L_E(\bigcup_{v\in V}\bigcup_{S\in
  M_v}S)$.  By construction of the link graph, the vertices of
$P_{\{M_v\}}$ are a union of pairs $S\in \bigcup_{v\in V}M_v$.  In
particular, for each $v\in V$, the intersection $P_{\{M_v\}}\cap J_v$
is a union of pairs $S\in M_v$.  Using EW2' we have
\begin{align*}
B(\vecx,\vecy,M)
&=\sum_{\vecx',\vecy'} \;\; \sumind \;\; \prod_{v\in V} B_v(\vecx'|_{J_v},\vecy'|_{J_v},M_v)\\
&=\sum_{\vecx',\vecy'} \;\; \sumind \;\; \prod_{v\in V} B_v((\vecx'\oplus \mathbf P_{\{M_v\}})|_{J_v},(\vecy'\oplus \mathbf P_{\{M_v\}})|_{J_v},M_v)\\
&=B(\vecx\oplus \mathbf S,\vecy\oplus \mathbf S,M).
\end{align*}
So $B$ witnesses that $\sem\phi$ is even-windable.
\end{proof}

\subsection{The near-assignments Markov chain}

Throughout this subsection fix an even-windable \sig{} $F\sigf J$ and
a partition $E$ of $J$ into pairs.  This can be thought of as a
circuit with one vertex.  We will define and study the
\emph{near-assignments Markov chain} for $(F,E)$.

Set $n=|J|$. For each $k\geq 0$ let $\Omega_k$ denote the set of
$k$-assignments of $J$ with respect to $E$ that satisfy $F(\vecx)>0$.
The state-space is $\Omega=\Omega_0\cup\Omega_2$.  The transitions are
Metropolis updates to states at distance two. More specifically, the transition
probability from $\vecx$ to $\vecy$ is defined to be
\begin{align*}
P(\vecx,\vecy)=
\begin{cases}
\frac{2}{n^2}\min(1,F(\vecy)/F(\vecx))&\text{ if $\dist{\vecx}{\vecy}=2$}\\
1-\frac{2}{n^2}\sum_{\vecy'\colon \dist{\vecx}{\vecy'}=2} \min(1,F(\vecy')/F(\vecx))&\text{ if $\vecy=\vecx$}\\
0&\text{ otherwise.}
\end{cases}
\end{align*}

(We will not consider the initial state to be part of the Markov chain
itself: the Markov chain is completely described by the matrix
$P\in\mathbb{R}^{\Omega\times\Omega}$.)  Define a probability distribution
$\pi$ on $\Omega$ by
$$ \pi(\vecx)=F(\vecx)\Bigg/\sum_{\vecy\in\Omega}F(\vecy)\qquad (\vecx\in\Omega).$$

By abuse of notation we will also denote $\sum_{\vecx\in X}\pi(\vecx)$
by $\pi(X)$ for subsets $X\subseteq\Omega$.  By adapting
the arguments of \cite{approxperm}, we will show:

\begin{theorem}\label{thm:namix}
For all $\vecx\in\Omega$ and all non-negative integers $t$, we have
$$\frac 1 2 \sum_{\vecy\in\Omega}|P^t(\vecx,\vecy)-\pi(\vecy)| \leq \frac 1 2 \pi(\vecx)^{-1/2}\exp(-t\pi(\Omega_0)^2/n^4)$$
\end{theorem}

Here $P^t$ denotes the $t$'th matrix power. The factor of $\frac 1 2$ is convention: the left hand side is called total variation distance.

We will use a congestion argument, with the following definitions.
A \defn{flow-path} is a directed path $\gamma$ in the transition
graph\footnote{the directed graph with vertex set $\Omega$ and an arc
  $(\vecx,\vecy)$ whenever $P(\vecx,\vecy)>0$.}, equipped with a
weight $\wt(\gamma)$, and also equipped with a label so that a set of
paths can include the same path more than once.
A \defn{flow} $\Gamma$ from $X\subseteq \Omega$ to $Y\subseteq \Omega$
is a set of flow-paths which each start in $X$ and end in $Y$,
satisfying
\[ \sum_{\text{paths $\gamma\in\Gamma$ from $x$ to $y$}}\wt(\gamma)=\pi(x)\pi(y)\qquad\text{ for all $x\in X$ and $y\in Y$.} \]

The \defn{congestion} of a flow $\Gamma$ is defined to be
\[ \rho(\Gamma)=\max_{\text{transitions $(\vecx,\vecy)$}} \frac{1}{\pi(\vecx)P(\vecx,\vecy)} \sum_{\text{$\gamma\in\Gamma$ with $\gamma\ni (\vecx,\vecy)$}} \wt(\gamma). \]
Be aware that this is not the same as the definition in
\cite{sinclair}: we are using a set of weighted paths rather than an
assignment of weights to paths. When applying the results of
\cite{sinclair} in the proof of Theorem \ref{thm:namix} we will need
to sum the total weight along each (unlabelled) path.

In the following arguments we will often use $k$-assignments (with respect to $E$)
of the form $\vecx\oplus\vecy$ for $\vecx\in\Omega_{k_1}$ and
$\vecy\in\Omega_{k_2}$.  Note that we do not require $F(\vecx)>0$ for
$k$-assignments $\vecx$, though we do require $F(\vecx)>0$ for
$\vecx\in\Omega_k$ ($\Omega_k$ is the set of ``satisfying''
$k$-assignments).  For any non-negative integer $k$, a
\defn{$k$-assignment-matching} (with respect to $E$) is a set $M$ of
disjoint pairs of $J$ such that exactly $k$ edges $\{i,j\}\in E$ have
exactly one endpoint, $i$ or $j$, in $\bigcup_{S\in M}S$.  In other
words, the characteristic vector of $\bigcup_{S\in M}S$ is a
$k$-assignment.

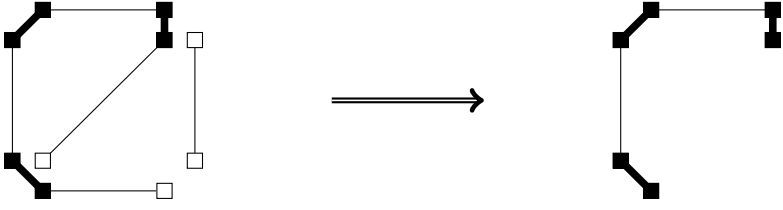
\begin{figure}
\begin{tikzpicture}


\foreach \x/\y in {0/0,0/2,2/0,2/2} {
}

\draw
(0.2,-0.2) -- (1.8,-0.2)
(2.2,0.2) -- (2.2,1.8)
(0.2,2.2) -- (1.8,2.2)
(-0.2,0.2) -- (-0.2,1.8)
(0.2,0.2) -- (1.8,1.8);

\draw[line width=0.1cm]
(-0.2,0.2) -- (0.2,-0.2)
(-0.2,1.8) -- (0.2,2.2)
(1.8,1.8) -- (1.8,2.2);

\begin{scope}[xshift=0cm,yshift=0cm]
\drawincg {0.2}{-0.2} ;
\drawinc {0.2}{0.2} ;
\drawincg {-0.2}{0.2} ;
\end{scope}

\begin{scope}[xshift=0cm,yshift=2cm]
\drawincg {0.2}{0.2} ;
\drawincg {-0.2}{-0.2} ;
\end{scope}

\begin{scope}[xshift=2cm,yshift=2cm]
\drawincg {-0.2}{-0.2} ;
\drawincg {-0.2}{0.2} ;
\drawinc {0.2}{-0.2} ;
\end{scope}

\begin{scope}[xshift=2cm,yshift=0cm]
\drawinc {0.2}{0.2} ;
\drawinc {-0.2}{-0.2} ;
\end{scope}

\draw[double, thick, ->] (4,1) -- (6,1);

\begin{scope}[xshift=8cm]

\draw
(0.2,2.2) -- (1.8,2.2)
(-0.2,0.2) -- (-0.2,1.8);

\draw[line width=0.1cm]
(-0.2,0.2) -- (0.2,-0.2)
(-0.2,1.8) -- (0.2,2.2)
(1.8,1.8) -- (1.8,2.2);

\begin{scope}[xshift=0cm,yshift=0cm]
\drawincg {0.2}{-0.2} ;
\drawincg {-0.2}{0.2} ;
\end{scope}

\begin{scope}[xshift=0cm,yshift=2cm]
\drawincg {0.2}{0.2} ;
\drawincg {-0.2}{-0.2} ;
\end{scope}

\begin{scope}[xshift=2cm,yshift=2cm]
\drawincg {-0.2}{-0.2} ;
\drawincg {-0.2}{0.2} ;
\end{scope}

\end{scope}

\end{tikzpicture}
\caption{The link graph $L_E(M)$ for some $2$-assignment-matching
  $M$. Squares are elements of $J$. Elements of $M$ are drawn as thick
  black lines. Elements $\{i,j\}\in E$ are drawn as thin
  lines.\label{fig:lg2}}
\end{figure}

Consider a $k$-assignment-matching $M$.  Recall the definition of the
link graph $L_E(M)$ given in Section \ref{sec:ewexpr}, which consists
of cycles and paths.  For any $i$ with $i\in \bigcup_{S\in M}S$, the
unique $j$ with $\{i,j\}\in E$ satisfies $j\in \bigcup_{S\in M}S$,
except for exactly $k$ values $i$. Thus $L_E(M)$ has precisely $k/2$
path components. See Figure \ref{fig:lg2}.

For all non-negative integers
$k$ define
$$Z_k=\sum_{\vecx\in\Omega_k}F(\vecx).$$
(This is $Z_k(\phi)$ if we consider $F$ as a one-vertex circuit $\phi$.)

\begin{lemma}\label{lem:z4}
$Z_0Z_4\leq Z_2Z_2$.
\end{lemma}
\begin{proof}
 We have
\begin{align*}
Z_0Z_4&=\sum_{\substack{\vecx\in\Omega_0\\\vecy\in\Omega_4}}F(\vecx)F(\vecy)
=\sum_{\substack{\vecx\in\Omega_0\\\vecy\in\Omega_4}}\sum_{M\in\Match{\vecx\oplus\vecy}}B(\vecx,\vecy,M).\\ \intertext{For
  each $4$-assignment-matching $M$, pick a path component of $L_E(M)$
  and let $H_M$ be the set of vertices of this component.  Let $B$ be
  a function witnessing that $F$ is even-windable. Each $H_M$ is a
  union of pairs in $M$ so by EW2',}
Z_0Z_4&=\sum_{\substack{\vecx\in\Omega_0\\\vecy\in\Omega_4}}\sum_{M\in\Match{\vecx\oplus\vecy}}B(\vecx\oplus\mathbf{H}_M,\vecy\oplus\mathbf{H}_M,M).\\ \intertext{
But $(\vecx\oplus\mathbf{H}_M,\vecy\oplus\mathbf{H}_M,M)$ determines $(\vecx,\vecy,M)$, and
  $\vecx\oplus\mathbf{H}_M,\vecy\oplus\mathbf{H}_M\in\Omega_2$. So}
Z_0Z_4&\leq\sum_{\substack{\vecx'\in\Omega_2\\\vecy'\in\Omega_2}}\sum_{M\in\Match{\vecx'\oplus\vecy'}}B(\vecx',\vecy',M)\\ &=\sum_{\substack{\vecx'\in\Omega_2\\\vecy'\in\Omega_2}}F(\vecx')F(\vecy')=Z_2Z_2.
\end{align*}
\end{proof}

\begin{lemma}\label{lem:flow0}
Assume $Z_0>0$. There is a flow $\Gamma_0$ from $\Omega_0$ to
$\Omega$, using flow-paths of length at most $n/2$, and with congestion
at most $\frac 1 2 n^3/\pi(\Omega_0)$.
\end{lemma}
\begin{proof}
We will first construct a ``winding'' enumeration $S(M,1),\dots,S(M,|M|)$ of
each $0$- or $2$-assignment-matching $M$.
The property we will need is that for each $0\leq k\leq |M|$, the
characteristic vector of $S(M,1)\cup \dots \cup S(M,k)$ is a $0$- or
$2$-assignment.

First define the final pair $T(M)=S(M,|M|)$ for all non-empty $0$- or
$2$-assignment-matchings $M$ as follows.  If $M$ is a non-empty
$0$-assignment-matching (so $L_E(M)$ consists of cycles), pick any
vertex $i\in L_E(M)$.  If $M$ is a non-empty $2$-assignment-matching,
pick an endpoint $i$ of the unique path component in $L_E(M)$.  In either case
let $j$ be the unique index with $\{i,j\}\in M$, and set
$T(M)=\{i,j\}$.  In any case
$L_E(M\setminus\{T(M)\})=L_E(M)\setminus\{i,j\}$ has at most one path
component.

So $M\setminus \{T(M)\}$ is a $0$- or $2$-assignment-matching.  By
induction on $|M|-k$ define $$S(M,k)=T(M\setminus
\{S(M,k+1),\dots,S(M,|M|)\}).$$ So $M\setminus
\{S(M,k+1),\dots,S(M,|M|)\}$ is always a $0$- or
$2$-assignment-matching. This completes the construction of $S(M,k)$.

Let $B$ be a function witnessing that $F$ is even-windable.
Let $\Gamma_0$ be the set consisting of a flow-path
$\gamma_{\vecx,\vecy,M}$ for each $\vecx\in\Omega_0$ and
$\vecy\in\Omega$ and $M\in\Match{\vecx\oplus\vecy}$, where
$\gamma_{\vecx,\vecy,M}$ is the flow-path
$$\vecx=\vecx\oplus \mathbf J_{M,0} \to \vecx\oplus \mathbf J_{M,1} \to \dots \to
\vecx\oplus \mathbf J_{M,|M|}=\vecy $$ equipped with weight
$B(\vecx,\vecy,M)/(Z_0+Z_2)^2$ and label $(\vecx,\vecy,M)$, where $\mathbf J_{M,k}$
denotes the characteristic vector of $S(M,1)\cup\dots\cup S(M,k)$.

$\Gamma_0$ is a flow from $\Omega_0$ to $\Omega$ because for all $\vecx\in \Omega_0$ and $\vecy\in\Omega_2$ we have
\[ \sum_{M\in\Match{\vecx\oplus\vecy}}B(\vecx,\vecy,M)/(Z_0+Z_2)^2=F(\vecx)F(\vecy)/(Z_0+Z_2)^2=\pi(\vecx)\pi(\vecy). \]

The congestion of $\Gamma_0$ is
\begin{align*}
\rho(\Gamma_0)&=\max_{\text{transitions $(\vecz,\vecz')$}} \frac{1}{\pi(\vecz)P(\vecz,\vecz')} \sum_{\text{$\gamma\in\Gamma_0$ with $(\vecz,\vecz')\in\gamma$}} \wt(\gamma) 
\intertext{But $\pi(\vecz)P(\vecz,\vecz')=\frac 2
{n^2}\min(\pi(\vecz),\pi(\vecz'))$, so}
\rho(\Gamma_o)&\leq \max_{\vecz\in\Omega} \frac{n^2}{2\cdot \pi(\vecz)} \sum_{\text{$\gamma\in\Gamma_0$ with $\vecz\in\gamma$}} \wt(\gamma) \\
&= \max_{\vecz\in \Omega} \frac{n^2}{2F(\vecz)(Z_0+Z_2)} \sum_{\substack{\vecx\in\Omega_0\\\vecy\in\Omega_2}} \;\;\sum_{\substack{M\in\Match{\vecx\oplus\vecy}\\\text{with }\vecz\in\gamma_{\vecx,\vecy,M}}} B(\vecx,\vecy,M)
\intertext{
In the last summation, $\vecz\in\gamma_{\vecx,\vecy,M}$ implies $\vecz=\vecx\oplus\mathbf J_{M,k}$ for some $k$,
so by EW2' we have $B(\vecx,\vecy,M)=B(\vecz,\vecz\oplus\vecw,M)$
where $\vecw=\vecx\oplus\vecy$. Thus,}
\rho(\Gamma_0)&\leq \max_{\vecz\in \Omega} \frac{n^2}{2F(\vecz)(Z_0+Z_2)} \sum_{\text{$0$- and $2$-assignments $\vecw$}} \;\;\; \sum_{\vecx\in\Omega_0}\sum_{\substack{M\in\Match{\vecw}\\\text{with }\vecz\in\gamma_{\vecx,\vecx\oplus\vecw,M}}} B(\vecz,\vecz\oplus\vecw,M)
\intertext{For each $(\vecz,\vecw,M)$ with
$M\in\Match{\vecw}$, the only values of $\vecx$ such that
$\vecz\in\gamma_{\vecx,\vecx\oplus\vecw,M}$ are the $|M|+1$ values $\vecz\oplus \mathbf
J_{M,0}, \dots, \vecz\oplus\mathbf J_{M,|M|}$. Thus,}
\rho(\Omega_0)&\leq \max_{\vecz\in \Omega} \frac{n^2}{2F(\vecz)(Z_0+Z_2)} \sum_{\text{$0$- and $2$-assignments $\vecw$}} (|M|+1) \sum_{\substack{M\in\Match{\vecw}}} B(\vecz,\vecz\oplus\vecw,M)\\
&= \max_{\vecz\in \Omega} \frac{n^2(n/2+1)}{2F(\vecz)(Z_0+Z_2)}  \sum_{\text{$0$- and $2$-assignments $\vecw$}} F(\vecz) F(\vecz\oplus\vecw)
\intertext{Using $\vecz\oplus\vecw\in\Omega_0\cup\Omega_2\cup\Omega_4$,}
\rho(\Omega_0)&\leq \frac{n^3}{2} \cdot \frac {Z_0+Z_2+Z_4}{Z_0+Z_2}.
\end{align*}

If $Z_2=0$ then by Lemma \ref{lem:z4} we also have $Z_4=0$, so the
congestion is at most $n^3/2$.  Otherwise by Lemma \ref{lem:z4} we
have $Z_4/Z_2\leq Z_2/Z_0$ and
\[ \frac{Z_0+Z_2+Z_4}{Z_0+Z_2} \leq 1 + \frac{Z_4}{Z_2} \leq 1 + \frac{Z_2}{Z_0} = 1 \big/ \frac{Z_0}{Z_0+Z_2} = 1/\pi(\Omega_0). \]
\end{proof}

\begin{lemma}\label{lem:flow}
Assume $Z_0>0$. There is a flow $\Gamma$ from $\Omega$ to $\Omega$,
using flow-paths of length at most $n$, and with congestion at most
$n^3/\pi(\Omega_0)^2$.
\end{lemma}
\begin{proof}
As in \cite{JSV}, we will randomly route through $\Omega_0$.

For each pair of flow-paths $g,g'$ starting at the same state $\vecy$, construct a
flow-path $\gamma(g,g')$ by appending $g'$ to the reverse of $g$, and
assigning a weight of $\wt(g)\wt(g')/\pi(\vecy)\pi(\Omega_0)$, and
assigning the label $(g,g')$.  Let $\Gamma_0$ be the set of flow-paths
given by Lemma \ref{lem:flow0}.  Let $\Gamma_0(\vecy,\vecx)$ denote
the set of flow-paths in $\Gamma_0$ starting at $\vecy$ and ending at
$\vecx$.  Let $\Gamma$ denote the set of flow-paths $\gamma(g,g')$
with $g\in\Gamma_0(\vecy,\vecx)$ and $g'\in\Gamma_0(\vecy,\vecz)$ for
some $\vecx,\vecz\in\Omega$ and $\vecy\in\Omega_0$.  Then $\Gamma$ is
a flow from $\Omega$ to $\Omega$ because for all $\vecx,\vecz\in
\Omega$ we have
\begin{align*}
\sum_{\vecy\in\Omega_0}
\sum_{\substack{g\in\Gamma_0(\vecy,\vecx)\\g'\in\Gamma_0(\vecy,\vecz)}}
\wt(g)\wt(g')/\pi(\vecy)\pi(\Omega_0)
&=\sum_{\vecy\in\Omega_0}\pi(\vecx)\pi(\vecy)\pi(\vecy)\pi(\vecz)/\pi(\vecy)\pi(\Omega_0)\\
&=\pi(\vecx)\pi(\vecz).
\end{align*}

Letting $(\vecw,\vecw')$ denote an arbitrary transition, the congestion of $\Gamma$
is
\begin{align*}
\rho(\Gamma)
&=\max_{(\vecw,\vecw')} \frac{1}{\pi(\vecw)P(\vecw,\vecw')}
\sum_{\substack{\vecx,\vecz\in\Omega\\\vecy\in\Omega_0}}
\sum_{\substack{g\in\Gamma_0(\vecy,\vecx)\\g'\in\Gamma_0(\vecy,\vecz)\\\text{such that $(\vecw,\vecw')\in \gamma(g,g')$}}}
 \frac{\wt(g)\wt(g')}{\pi(\vecy)\pi(\Omega_0)}.
 \intertext{By symmetry,}
\rho(\Gamma)&=2\max_{(\vecw,\vecw')} \frac{1}{\pi(\vecw)P(\vecw,\vecw')}
\sum_{\substack{\vecx,\vecz\in\Omega\\\vecy\in\Omega_0}}
\sum_{\substack{g\in\Gamma_0(\vecy,\vecx)\\g'\in\Gamma_0(\vecy,\vecz)\\\text{such that $(\vecw,\vecw')\in g'$}}}
 \frac{\wt(g)\wt(g')}{\pi(\vecy)\pi(\Omega_0)}\\
&=2\max_{(\vecw,\vecw')} \frac{1}{\pi(\vecw)P(\vecw,\vecw')}
\sum_{\substack{\vecx,\vecz\in\Omega\\\vecy\in\Omega_0}}
\sum_{\substack{g'\in\Gamma_0(\vecy,\vecz)\\ \text{such that $(\vecw,\vecw')\in g'$}}}
 \frac{\pi(\vecx)\pi(\vecy)\wt(g')}{\pi(\vecy)\pi(\Omega_0)}\\
&=2\max_{(\vecw,\vecw')} \frac{1}{\pi(\vecw)P(\vecw,\vecw')}
\sum_{\substack{\vecz\in\Omega\\\vecy\in\Omega_0}}
\sum_{\substack{g'\in\Gamma_0(\vecy,\vecz)\\ \text{such that $(\vecw,\vecw')\in g'$}}}\wt(g')/\pi(\Omega_0)\\
&=2\rho(\Gamma_0)/\pi(\Omega_0)\\
&\leq n^3/\pi(\Omega_0)^2
\end{align*}
by Lemma \ref{lem:flow0}.
\end{proof}

The remaining task is to relate the congestion to Markov chain mixing.

\begin{reptheorem}{thm:namix}
For all $\vecx\in\Omega$ and all non-negative integers $t$, we have
$$\frac 1 2 \sum_{\vecy\in\Omega}|P^t(\vecx,\vecy)-\pi(\vecy)| \leq \frac 1 2 \pi(\vecx)^{-1/2}\exp(-t\pi(\Omega_0)^2/n^4).$$
\end{reptheorem}
\begin{proof}
The transition matrix $P$ is reversible relative to $\pi$: it obeys the detailed balance condition
\[\pi(\vecy)P(\vecy,\vecz)=\pi(\vecz)P(\vecz,\vecy)\text{ for all $\vecy,\vecz\in\Omega$.}\]
We have
\begin{align}\label{eq:selfloop}
P(\vecx,\vecx) \geq 1-\frac{2}{n^2}\binom{n}{2} \geq 1/n\qquad\text{ for all $\vecx\in\Omega$.}
\end{align}
In particular, the Markov chain is aperiodic.  Also, by Lemma
\ref{lem:flow} there exists a flow $\Gamma$ from $\Omega$ to $\Omega$,
which implies that that the Markov chain is connected. This allows us
to use the results from \cite{sinclair} and \cite{strook}.
 $P$ has eigenvalues
$$1=\lambda_0>\lambda_1\geq \dots \lambda_{|\Omega|-1}\geq -1.$$
By setting $f(\gamma)$ in \cite[Corollary 6']{sinclair} to be the sum of
the weight of flow-paths in $\Gamma$ whose underlying directed path is
$\gamma$, we have
\begin{align*}
\lambda_1  &\leq 1 - \frac{1}{\rho(\Gamma)n}\\
&\leq 1-\pi(\Omega_0)^2/n^4.
\end{align*}
using Lemma \ref{lem:flow}. By \eqref{eq:selfloop} and equation 1 of \cite{lazy} we have
$$-\lambda_{|\Omega|-1}\leq 1-2/n \leq \lambda_1.$$
By \cite[Proposition 3]{strook},
\begin{align*}
\frac 1 2 \sum_{\vecy\in\Omega}|P^t(\vecx,\vecy)-\pi(\vecy)|
&\leq \frac 1 2 \sqrt{\frac{1-\pi(\vecx)}{\pi(\vecx)}}\max(\lambda_1,-\lambda_{|\Omega|-1})^t\\
&\leq \frac 1 2 \pi(\vecx)^{-1/2}\exp(-t\pi(\Omega_0)^2/n^4).
\end{align*}
\end{proof}

\section{Windable functions}\label{sec:windable}

In this section we extend the analysis of even-windable functions to
windable functions. The definition of windability is a natural
extension of even-windability, but turns out not to give much extra
generality.

For all $F\sigf J$ define $F_\oplus\sigf{1+J}$ by
$$F_{\mathrm{\oplus}}(p;\vecx)=
\begin{cases}
F(\vecx)&\text{ if $p+\sum_{i\in J}x_i$ is even}\\
0&\text{ otherwise}
\end{cases}\qquad (p\in\{0,1\}, \vecx\in\{0,1\}^J).$$

\begin{lemma}\label{lem:windchar}
$F\sigf J$ is windable if and only if $F_\oplus$ is even-windable.
\end{lemma}
\begin{proof}
($\Rightarrow$) Pick an ordering of $J$.  Consider a partition $M$ of
  a subset $I\subseteq J$ into singletons $\{a_1\},\dots,\{a_k\}$ and
  pairs $S_1,\dots,S_\ell$. Define $\mu(M)$, when $|I|$ is even, to be
  the union of $\{S_1,\dots,S_\ell\}$ with a partition (depending only on $M$) of
  $\{a_1,\dots,a_k\}$ into pairs.  Define $\mu(M)$, when $|I|$ is
  odd, to be the union of $\{S_1,\dots,S_\ell\}$ with a partition of
  $\{1,a_1,\dots,a_k\}$ into pairs. Let $B$ be a witness that $F$ is
  windable. For all $(p;\vecx),(q;\vecy)\in\{0,1\}^{1+J}$
and all $M\in\Match{(p;\vecx)\oplus(q;\vecy)}$, define
\begin{align*}
B'((p;\vecx),(q;\vecy),M)=
\begin{cases}
\sum\limits_{M'\colon \mu(M')=M} B(\vecx,\vecy,M')&\text{ if
$p+\sum_{i\in J}x_i$ and $q+\sum_{i\in J}y_i$ are even}\\
0&\text{ otherwise}\\
\end{cases}
\end{align*}
For all $S\in M=\mu(M')$, if we let $S'=S\setminus \{1\}$ then
$B(\vecx\oplus\mathbf{S'}, \vecy\oplus\mathbf
{S'},M')=B(\vecx,\vecy,M')$. So $B'$ witnesses that $F_\oplus$ is
even-windable.

($\Leftarrow$) For all sets $M$ of disjoint pairs of $1+J$ define
$\nu(M)$ to be $\{S\setminus\{1\}\mid S\in M\}$. Let $B$ be a witness
that $F_\oplus$ is windable. For all $\vecx,\vecy\in\{0,1\}^J$ and all
$M\in\Matchp{\vecx\oplus\vecy}$, define
\begin{align*}
B'(\vecx,\vecy,M)=\sum_{p,q=0}^1 \sum_{M'\colon \nu(M')=M} B((p;\vecx),(q;\vecy),M')
\end{align*}

For all $S\in M=\nu(M')$, let $S'=S$ if $|S|=2$ and $S'=S\cup\{1\}$
otherwise.  Then $B(\vecx\oplus\mathbf{S'}, \vecy\oplus\mathbf
{S'},M')=B(\vecx,\vecy,M')$. So $B'$ witnesses that $F$ is windable.
\end{proof}

\begin{lemma}\label{lem:windexpr}
Let $\phi$ be a circuit using only \sigs{} that are windable.
The \sig{} of $\phi$ is windable.
\end{lemma}
\begin{proof}
Replace each constraint $F_v$ by $(F_v)_\oplus$, rename the new
incidences $p_v$, $v\in V(\phi)$, and add a constraint
$\sigEven_{1+P}$ where $P=\{p_v\mid v\in V(\phi)\}$.  This produces a
circuit $\phi_\oplus$ with $\sem{\phi_\oplus}=\sem{\phi}_\oplus$.  By
Lemmas \ref{lem:windchar}, \ref{lem:evenwind_expr}, and
\ref{lem:evenwind_paritynae} we find that $\sem{\phi_\oplus}$ is
even-windable.  So by Lemma~\ref{lem:windchar} again, $\sem{\phi}$ is
windable.
\end{proof}

\begin{lemma}\label{lem:flipdecomp}
For any $J$, the \sigs{} $\sigEven_J$, $\sigOdd_J$, and $\sigNAE_J$ are windable.
\end{lemma}
\begin{proof}
By Lemma \ref{lem:evenoddeven} there is a witness $B$ that
$\sigEven_J$ is even-windable. Extending $B$ by setting
$B(\vecx,\vecy,M)=0$ for all
$M\in\Matchp{\vecx\oplus\vecy}\setminus\Match{\vecx\oplus\vecy}$ we
get a witness $B'$ that $\sigEven_J$ is windable.  Similarly,
$\sigOdd_J$ is even-windable by Lemma \ref{lem:evenoddeven}, and it is
therefore windable.

For $\sigNAE_j$, by Lemma \ref{lem:windchar} it suffices to show that
the \sig{} $(\sigNAE_J)_{\oplus}$ is even-windable.  
Let $I\subseteq 1+J$, let
$\vecp\in\{0,1\}^I$, let $K=(1+J)\setminus I$ and let $G\sigf K$ be
the pinning of $(\sigNAE_J)_{\oplus}$ by $\vecp$.  We wish to show
that $G\overline G$ has a 2-decomposition.  If $|K|$ is odd then
$G\overline G$ is identically zero so has a 2-decomposition. We can
therefore assume that $|K|$ is even.

Let $c\in\{0,1\}$ be equal to $|J|$ modulo $2$.
$(\sigNAE_J)_{\oplus}$ is the \sig{} corresponding to the relation
$\relEven_{1+J}\setminus X$ where $X$ consists of two configurations
at distance $|J|+c$. Specifically, $X$ contains the all-zeros
configurations of $1+J$, and also contains $(c;\vecx)$ where $\vecx$
is the all-ones configuration.
We first argue that in all cases, $G\overline G$ is either $\sigEven_K$ or $\sigOdd_K$, or a flip of $\sigEvenNAE_K$.

If $\sum_{i\in I}p_i$ is odd, then $G$ takes the value $1$ precisely
on $\relOdd_{K}\setminus X'$ where $X'$ consists of at most one
configuration $\vecx\in\relOdd_K$.  If $X'=\emptyset$ then $G\overline
G=\sigOdd_K$.  If $X'=\{\vecx\}$ then $G\overline G$ is the flip of
$\sigEvenNAE_K$ by $\vecx$.

If $\sum_{i\in I}p_i$ is even, then $G$ takes the value $1$ precisely
on $\relEven_{K}\setminus X'$ where $X'$ consists of at most two
configurations in $\relEven_K$.  If $|X'|\leq 1$ we are done by the
same argument as the previous paragraph: $G\overline G$ is either
$\sigEven_K$ or a flip of $\sigEvenNAE_K$.  If $|X'|=2$ then $X'$
consists of two configurations $\vecx,\vecy$ with
$\dist{\vecx}{\vecy}=|J|+c$. But $|J|+c\leq |K|\leq |J|+1$, and $|K|$
and $|J|+c$ are both even, so $|K|=|J|+c$. Thus
$\vecy=\overline{\vecx}$, and again $G\overline G$ is a flip of
$\sigEvenNAE_K$.

By Lemma \ref{lem:evenoddeven} the \sigs{} $\sigEven_K$ and
$\sigOdd_K$ have 2-decompositions.  So we only need to check the last
case where $G\overline G$ is a flip, by $\vecz\in\{0,1\}^K$ say, of
$\sigEvenNAE_K$. Let $D$ be a 2-decomposition of $\sigEvenNAE_K$ given
by Lemma \ref{lem:evenwind_paritynae}. Define
$D'(\vecx,M)=D(\vecx\oplus\vecz,M)$ for all $\vecx\in\{0,1\}^K$ and
for all partitions $M$ of $K$ into pairs.  For all
$\vecx\in\{0,1\}^K$ we have $G\overline
G(\vecx)=\sigEvenNAE_K(\vecx\oplus\vecz)=\sum_M D(\vecx\oplus\vecz,M)=\sum_M
D'(\vecx,M)$, where $M$ ranges over partitions of $K$ into pairs.  So
$D'$ is a 2-decomposition of $G\overline G$.
\end{proof}

\section{Strictly terraced functions}\label{sec:strterr}

\subsection{Idea}

To apply Theorem \ref{thm:namix} to Holant problems, the challenge is
to find a class of circuits for which the ratio of the weight of
$2$-assignments to the weight of $0$-assignments is polynomially
bounded in the size of the Holant instance.

The weight of $2$-assignments of a closed circuit can be written in
terms of the \sigs{} obtained by breaking two edges; the challenge
then reduces to trying to find a bound on the ratios
$F(\vecx)/F(\vecy)$ between the values in these \sigs{}, when
$F(\vecy)\neq 0$.

It is instructive to consider multiplication of two-by-two matrices.
  To see the relationship between multiplication of matrices and
circuits (in the form of read-twice pps-formulas), for matrices $M$
with rows and columns indexed by $\{0,1\}$, define $F_M\sigf 2$ by
$F_M(i,j)=M_{i,j}$; then $F_{MN}(i,k)=\sum_j F_M(i,j)F_N(j,k)$.

Matrix multiplication can produce exponentially-large ratios: for any
$x,y>0$, we have
$$\begin{pmatrix}x&y\\0&1\end{pmatrix}^n
=\begin{pmatrix}x^n&y(x^{n-1}+\dots+1)\\0&1\end{pmatrix}
$$
and $x^n/1$ is exponentially large if $x>1$.

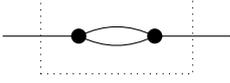
\begin{figure}
\begin{tikzpicture}

\draw[dotted] (-0.5,-0.5) rectangle (1.5,0.5);
\draw (-1,0) -- (0,0);
\draw (1,0) -- (2,0);
\draw (0,0) to[out=25,in=155] (1,0);
\draw (0,0) to[out=-25,in=-155] (1,0);
\drawvert{0}{0};
\drawvert{1}{0};

\end{tikzpicture}
\caption{A circuit with \sig{} $F$ with $F(0,0)=2$ and $F(1,1)=1$ and
  $F(0,1)=F(1,0)=0$. Vertices represent ``exact-one'' constraints
  $\{(1,0,0),(0,1,0),(0,0,1)\}$.\label{fig:broder}}
\end{figure}

In fact, the matrix
  $\left(\begin{smallmatrix} 2 & 0 \\ 0 &
    1 \end{smallmatrix}\right)$ corresponds
  to the circuit depicted in Figure \ref{fig:broder} using
  ``exact-one'' constraints $\{(1,0,0),(0,1,0),(0,0,1)\}$, which can
  be used to construct counterexamples to the bound \eqref{eq:npm} on
  nearly perfect matchings \cite{Bro86}.

We might guess that exponentially-large ratios can only be produced by
matrix multiplication when the zero entry in the matrix is surrounded
by values that are different. And indeed this property of being
``strictly terraced'' turns out to give some control over ratios. For
strictly terraced functions, the worst ratio in a \sig{} is bounded by
the sum of the worst ratios that can be obtained by mixing the
individual functions with parity relations.

\subsection{Definitions}

A function $F\sigf J$ is
\defn{strictly terraced} if
\[ F(\vecx)=0 \implies F(\vecx\oplus\eee i)=F(\vecx\oplus\eee j)\qquad\text{ for all $\vecx\in\{0,1\}^J$ and all $i,j\in J$}.\]

For all \sigs{} $F$ that are not identically zero, a
\defn{parity-\sig{} of $F$} is a constant multiple of the \sig{} of a
circuit using one $F$ constraint and such that all other constraints
are parity relations.  Define
\begin{align*}
\Prat(F)=\max \left\{ \frac{F'(0)}{F'(1)} \middle\vert \text{\parbox{2.7in}{\centering{$F'\colon\{0,1\}\to\Qnonneg$ is a parity-\sig{} of $F$ with $F'(1)>0$}}} \right\}.
\end{align*}
We extend $\Prat$ to all \sigs{} $F$ by setting $\Prat(F)=0$
if $F$ is identically zero.

We can show that $\Prat$ is well-defined using the following
operation.  For any circuit $\phi$ and any internal edge $e\in E^\phi$
between incidences $i_u\in J_u$ and $i_v\in J_v$, with $u,v\in V$ (not
necessarily distinct), define the \defn{contraction of $\phi$ by $e$}
to be the circuit $\phi'$ obtained by replacing $u$ and $v$ by a
vertex $w$ with incidences $(J_u\cup J_v)\setminus \{i_u,i_v\}$ and
equipping $w$ with the \sig{} of the circuit with constraints $F_u$
and $F_v$, edge $e$, and external edges $(J_u\cup J_v)\setminus
\{i_u,i_v\}$.

\begin{lemma}\label{lem:contrparity}
Let $\phi$ be a connected circuit whose constraints are all parity
constraints.  The \sig{} of $\phi$ is a constant multiple of a parity
constraint.
\end{lemma}
\begin{proof}
By induction on the number of edges of $\phi$, it suffices to show
that contracting a single edge of $\phi$ leaves only parity
constraints (up to multiplication by constants).

Consider the case that the edge $\{i,j\}$ goes between distinct
vertices, which are equipped with a $\sigEven_{\{i\}\cup I}$
constraint and a $\sigEven_{\{j\}\cup J}$ constraint.  Then
contraction gives a copy of $\sigEven_{I\cup
  J}$, because $\sigEven_{I\cup J}(\vecx,\vecy)=\sum_t\sigEven_{1+I}(t;\vecx)\sigEven_{1+J}(t;\vecy)$ for all $\vecx\in\{0,1\}^I$ and all $\vecy\in\{0,1\}^J$.
Similarly $\sigOdd_{\{i\}\cup I}$ and $\sigOdd_{\{j\}\cup J}$ produce
$\sigEven_{I\cup J}$, while $\sigEven_{\{i\}\cup I}$ and
$\sigOdd_{\{j\}\cup J}$ produce $\sigOdd_{I\cup J}$.

If the edge $\{i,j\}$ is a loop on a vertex with constraint
$\sigEven_{\{i,j\}\cup J}$, contracting $\{i,j\}$ produces
$2\sigEven_{J}$. Similarly $\sigEven_{\{i,j\}\cup J}$
produces $2\sigOdd_{J}$.
\end{proof}

Note that contracting an edge does not affect the \sig{} of a circuit.
By contracting edges between parity relations, the circuits appearing
in the definition of a parity-\sig{} can be rewritten not to use any
edges except external edges and edges incident to the $F$
constraint. For fixed $F$ there are therefore a finite number of
equivalence classes of parity-\sigs{} $F'\colon\{0,1\}\to\Qnonneg$
with $F'(1)>0$, under the equivalence relation of multiplication by
constants.  Thus the maximum in the definition of $\Prat(F)$ is taken
over a finite set, which can be seen to be non-empty if $F$ is not
identically zero (if $F(\vecx)>0$ for some $\vecx\in\relEven_J$ then
the function $F'\sigf 1$ defined by
$F'(t)=\sum_{\vecx\in\{0,1\}^J}\sigOdd_1(t)F(\vecx)\sigEven_J(\vecx)$
satisfies $F'(1)>0$, and if $F(\vecx)>0$ for some $\vecx\in\relOdd_J$
then the function $F'\sigf 1$ defined by
$F'(t)=\sum_{\vecx\in\{0,1\}^J}\sigOdd_1(t)F(\vecx)\sigOdd_J(\vecx)$
satisfies $F'(1)>0$).

Note that if $G$ is a parity-\sig{} of $F$, then $\Prat(G)\leq\Prat(F)$.
In particular if $G$ is a pinning of $F$ then $\Prat(G)\leq\Prat(F)$.
Also, since the disequality relation $\relOdd_2=\{(0,1),(1,0)\}$ is a parity relation,
it is not important that we took $F'(0)/F'(1)$ rather
than $F'(1)/F'(0)$ in the definition of $\Prat$.

\subsection{Examples}

A relation $R\subseteq\{0,1\}^J$ is \defn{coindependent} if for all
$\vecx\in\{0,1\}^J\setminus R$ we have $\vecx\oplus\eee i\in R$ for
all indices $i$. For example, the disequality relation
$\{(0,1),(1,0)\}$ is coindependent.
Any coindependent relation
$R$ gives an example $\mathbf R$ of a strictly terraced \sig{}.

\begin{lemma}\label{lem:pratex}
For all finite sets $J$, the functions $\sigEven_J$, $\sigOdd_J$ and
$\sigNAE_J$ are strictly terraced.  Also,
$\Prat(\sigEven_J)=\Prat(\sigOdd_J)=0$, and $\Prat(\sigNAE_J)\leq 3$.
\end{lemma}
\begin{proof}
The first statement follows from the fact that the corresponding
relations are coindependent.  To show
$\Prat(\sigEven_J)=\Prat(\sigOdd_J)=0$, note that by Lemma
\ref{lem:contrparity} a parity-\sig{} of a parity relation must be even.

Now we will show that $\Prat(\sigNAE_J)\leq 3$.  Consider a connected
circuit $\phi$ with one external edge, such that $\phi$ uses one
$\sigNAE_J$ constraint, and all other constraints are parity
relations, with no internal edges between parity relations (this is
without loss of generality, because we can contract any such edge).
Assume that $\sem{\phi}(0)$ and $\sem{\phi}(1)$ are non-zero.  We will
show that $\sem{\phi}(0)\leq 3\sem{\phi}(1)$.

We can write
$$\sem{\phi}(t)=\sum_{\vecx\in\relNAE_J} \mathbf R(t;\vecx)$$ where
$R$ is an affine subspace of $\mathrm{GF}(2)^{1+J}$.  Since $\sem{\phi}(0)$ and
$\sem{\phi}(1)$ are non-zero, the sets $R_0=\{\vecx\mid (0;\vecx)\in
R\}$ and $R_1=\{\vecx\mid (1;\vecx)\in R\}$ are non-empty.  Since $R$
is an affine subspace, $|R_0|=|R_1|$, so
$$\sem{\phi}(0) \leq |R_0| = |R_1| \leq \sem{\phi}(1)+2 \leq 3\sem{\phi}(1).$$
\end{proof}

\subsection{Properties}
An important property we will use is that a strictly terraced function $F$
is either identically zero or its support $\{\vecx\mid F(\vecx)>0\}$ is coindependent.  (If
$F(\vecx)=0$ and $F(\vecy)>0$ for some $\vecy$, pick such a $\vecy$
with $d=\dist{\vecx}{\vecy}$ minimal. If $d>1$, there are distinct
indices $i,j$ such that $x_i\neq y_i$ and $x_j\neq y_j$, so
$F(\vecy\oplus \eee i)=F(\vecy\oplus \eee i\oplus\eee j)=0$ by
minimality of $\dist{\vecx}{\vecy}$, which means $F$ is not strictly
terraced: $F(\vecy\oplus\eee i)=0$ but $F((\vecy\oplus\eee i)\oplus \eee i)\neq F((\vecy\oplus\eee i)\oplus\eee j)$.)  

The Cartesian product of coindependent relations is in general not
coindependent, for example $\{(0,1),(1,0)\}\times\{(0,1),(1,0)\}$ is
not coindependent (set $\vecx=(0,0,0,0)$ and $i=1$). Thus the class of
strictly terraced functions is not closed under taking \sigs{} of
disconnected circuits.

\begin{lemma}\label{lem:st_expr}
Let $\phi$ be a connected circuit using strictly terraced \sigs{}.
Then $\sem{\phi}$ is strictly terraced.
\end{lemma}
\begin{proof}
We will argue by induction on the number of internal edges of $\phi$.
If there are no internal edges, then $\phi$ consists of a single
constraint using a strictly terraced function $F$, and
$\sem{\phi}=F$. Otherwise, pick an internal edge $e$.  We wish to argue
that the function created by contracting $e$ is strictly terraced.
There are two cases.

(i) $e$ is loop on a vertex $v$.

Let $F\sigf {2+J}$ be a copy of $F_v$, indexed so that the ends of $e$
become enumerated indices.  We wish to show that the function
$H\sigf J$ defined by
$$H(\vecx)=\sum_{t=0}^1 F(t,t;\vecx)\qquad (\vecx\in\{0,1\}^J)$$ is
strictly terraced.  Consider $\vecx\in\{0,1\}^J$ satisfying
$H(\vecx)=0$ and let $i,j\in J$.  Since $F$ is strictly terraced and
$F(0,0;\vecx)=F(1,1;\vecx)=0$, we have $F(0,0;\vecx\oplus\eee
i)=F(0,0;\vecx\oplus\eee j)$ and $F(1,1;\vecx\oplus\eee
i)=F(1,1,\vecx\oplus\eee j)$.  Hence $H(\vecx\oplus\eee
i)=H(\vecx\oplus\eee j)$.

(ii) $e$ is incident to distinct vertices $u$ and $v$.

Let $F\sigf {1+I}$ and $G\sigf {1+J}$ be copies of $F_u$ and $F_v$
respectively, reindexed so that the ends of $e$ become the enumerated
indices (and with $I$ and $J$ disjoint).  We wish to show that the
function $H\sigf {I\cup J}$ defined by
$$H(\vecx,\vecy)=\sum_{t=0}^1 F(t;\vecx)G(t;\vecy)\qquad
(\vecx\in\{0,1\}^I, \vecy\in\{0,1\}^J)$$ is strictly terraced.  If $F$
or $G$ is identically zero then $H$ is identically zero and therefore
strictly terraced.

Otherwise consider $\vecx\in\{0,1\}^I$ and $\vecy\in\{0,1\}^J$
satisfying $H(\vecx,\vecy)=0$.  Since $F$ and $G$ have coindependent
support and $F(0;\vecx)G(0;\vecy)+F(1;\vecx)G(1;\vecy)=0$, there
exists $t\in\{0,1\}$ such that $F(t;\vecx)=G(1-t;\vecy)=0$ and
$F(1-t;\vecx),G(t;\vecy)> 0$.  For all $i\in I$ we have
$$H(\vecx\oplus\eee i,\vecy)=F(t;\vecx\oplus\eee
i)G(t;\vecy)=F(1-t;\vecx)G(t;\vecy).$$ Similarly for $i\in J$ we have
$$H(\vecx,\vecy\oplus\eee i)=F(1-t;\vecx)G(1-t;\vecy\oplus\eee i)=F(1-t;\vecx)G(t;\vecy).$$
Therefore for all $i,j\in I\cup J$ we have
$H((\vecx,\vecy)\oplus\eee i)=H((\vecx,\vecy)\oplus\eee j)$. 
\end{proof}

The following calculations bound ratios
produced by certain circuits.

\begin{lemma}\label{lem:pratrtfhelper}
Let $F\colon\{0,1\}^{1+J}$ and $G\sigf J$. Define $H(0),H(1)$ by
$$H(t)=\sum_{\vecx\in\{0,1\}^J}F(t;\vecx)G(\vecx).$$
Assume that $F$ and $G$ are strictly terraced and that $H(1)>0$. Then $$H(0)\leq (\Prat(F)+\Prat(G))H(1).$$
\end{lemma}
\begin{proof}
We will use induction on $|J|$. For the base case $J=\emptyset$ we
have $H(0)\leq \Prat(F)H(1)$ by definition of $\Prat(F)$.  So assume
that $J$ is non-empty.

For each $i\in J$ and each $c\in\{0,1\}$ define
$F_{i,c}$ to be the pinning of $F$ by taking $i$ to $c$,
and similarly define $G_{i,c}$ to be the pinning of $g$ by taking $i$ to $c$,
and define
$$H_{i,c}(t)=\sum_{\vecx\in\{0,1\}^{J\setminus\{i\}}} F_{i,c}(t,\vecx)G_{i,c}(\vecx)\qquad (t\in\{0,1\}).$$

Since pinnings are parity-\sigs{}, $\Prat(F_{i,c})\leq \Prat(F)$
and $\Prat(G_{i,c})\leq \Prat(G)$.
If there exists $i\in J$ such that $H_{i,0}(1)$ and $H_{i,1}(1)$ are non-zero, then by
the induction hypothesis we have
$$H(0)=H_{i,0}(0)+H_{i,1}(0)\leq
(\Prat(F)+\Prat(G))(H_{i,0}(1)+H_{i,1}(1))=(\Prat(F)+\Prat(G))H(1).$$
Taking a choice for each $i$, we may assume that there exists
$y\in\{0,1\}^J$ such that for all $i\in J$ we have $H_{i,1-y_i}(1)=0$.
So for each $i\in I$ the sets $R=\{\vecx\mid F_{i,1-y_i}(1;\vecx)>0\}$
and $S=\{\vecx\mid G_{i,1-y_i}(\vecx)>0\}$ are disjoint.  $R$ and $S$
are pinnings of coindependent relations, so they are coindependent.
For all $\vecx\in R$ we have $\vecx\notin S$, so $\vecx\oplus\eee i\in
S$ for any $i$, and $\vecx\oplus\eee i\notin R$. Repeatingly this, we
find that $R$ consists of the configurations at even distance from
$\vecx$, and $S$ consists of the configurations at odd distance from
$\vecx$.  In other words, for each $i\in J$ there exists
$c_i\in\{0,1\}$ such that
\begin{equation}\label{eq:parityfg}
\begin{aligned}
F(1,\vecx)>0 &\iff c_i+\sum_{j\in J} x_j\text{ is even, and}\\
G(\vecx)>0 &\iff c_i+\sum_{j\in J} x_j\text{ is odd}.
\end{aligned}
\qquad (\vecx\in\{0,1\}^J, x_i\neq y_i)
\end{equation}

For any $i,j\in J$ there is some choice of $\vecx\in\{0,1\}^J$ with
$x_i\neq y_i$ and $x_j\neq y_j$, so $c_i=c_j$. Thus there is a single
choice of $c$ such that \eqref{eq:parityfg} holds for all $i$ taking
$c_i=c$:

\begin{equation}\label{eq:parityfgstrong}
\begin{aligned}
F(1,\vecx)>0 &\iff c+\sum_{j\in J} x_j\text{ is even, and}\\
G(\vecx)>0 &\iff c+\sum_{j\in J} x_j\text{ is odd}.
\end{aligned}
\qquad (\vecx\in\{0,1\}^J\setminus\{\vecy\})
\end{equation}

For any $\vecx\in\{0,1\}^J$ with $c+\sum_{i\in J} x_i$ even,
and any distinct $i,j\in J$, we have $F(1;\vecx)=F(1;\vecx\oplus\eee i\oplus \eee j)$
because $F$ is strictly terraced and either $F(1;\vecx\oplus\eee i)$ or $F(1;\vecx\oplus\eee j)$ is zero.
Similarly for any $\vecx\in\{0,1\}^J$ with $c+\sum_{i\in J} x_i$ odd,
and any distinct $i,j\in J$, we have $G(\vecx)=G(\vecx\oplus\eee i\oplus\eee j)$.
This implies that there are constants $\lambda,\mu>0$ such that
\begin{equation}\label{eq:parityfgval}
\begin{aligned}
F(1;\vecx)&=\lambda\text{ if }c+\sum_{i\in J}x_i\text{ is even, and}\\
G(\vecx)&=\mu\text{ if }c+\sum_{i\in J} x_i\text{ is odd.}
\end{aligned}
\qquad (\vecx\in\{0,1\}^J).
\end{equation}

If $c+\sum_{i\in J} y_i$ is odd then by \eqref{eq:parityfgval} and
\eqref{eq:parityfgstrong}, $G$ is $\mu\sigEven_J$ (if $c=1$) or
$\mu\sigOdd_J$ (if $c=0$).  So $G$ is a constant multiple of a parity
relation.
Considering $H$ as the \sig{} of a circuit with
constraints $F$ and $G$, we get
$H(0)\leq \Prat(F)H(1)$ by definition of $\Prat$.  We may
therefore assume that $c+\sum_{i\in J} y_i$ is even.

Define $F'(0),F'(1),G'(0),G'(1)$ by
\begin{align*}
F'(t)&=F(t;\vecy)\\
G'(t)&=\sum_{\vecx\in\{0,1\}^J}\sigOdd_{2+J}(t,c;\vecx)G(\vecx)
\intertext{so}
F'(0)&=F(0;\vecy)\\
F'(1)&=F(1;\vecy)=\lambda\\
G'(0)&=\mu 2^{|J|-1}\\
G'(1)&=G(\vecy)
\end{align*}
Since $F'$ is a parity-\sig{} of $F$ we have $F(0;\vecy)/\lambda\leq
\Prat(F)$. Since $G'$ is a parity-\sig{} over
$G$ we have $\mu 2^{|J|-1}/G(\vecy)\leq \Prat(G)$.
For all $\vecx\in\{0,1\}^J$ with $c+\sum_{i\in J} x_i$ odd,
we have $F(1;\vecx)=0$ and $F(1;\vecx\oplus\eee i)=\lambda$ (for any $i\in J$)
and therefore $F(0;\vecx)=\lambda$
because $F$ is strictly terraced.
So
\begin{align*}
\frac{H(0)}{H(1)}=\frac
{F(0,\vecy)G(\vecy) + \lambda \mu 2^{|J|-1}}
{\lambda G(\vecy)} \leq \Prat(F)+\Prat(G).
\end{align*}
\end{proof}

\begin{lemma}\label{lem:pratbd}
Let $\phi$ be a circuit using strictly terraced \sigs{}. Then
\begin{align*}
\Prat(\sem{\phi})\leq \sum_{v\in V^\phi} \Prat(F^\phi_v).
\end{align*}
\end{lemma}
\begin{proof}
We will argue by induction on the number $k$ of constraints that are not parity relations.
The cases $k=0$ and $k=1$ follow from the definition of $\Prat$.
Components of a circuit not connected to the external edges
simply contribute a constant factor to the \sig{}.
So for a given $k$, it suffices to show that $\sem{\phi(0)}/\sem{\phi(1)}\leq \sum_{v\in V^\phi} \Prat(F^\phi_v)$ whenever:
\begin{itemize}
\item $\phi$ is a connected circuit with one external edge, with $\sem{\phi(1)}>0$, and
\item $\phi$ uses strictly terraced \sigs{}, at most $k$ of which are not parity relations.
\end{itemize}

For the $k=2$ case, if there is a loop on a vertex $v$, contract it.
This changes the \sig{} $F_v$, but the resulting \sig{} is a
parity-\sig{} of $F_v$, so this process does not increase $\sum_{v\in
  V}\Prat(F_v)$.  And $F_v$ is still strictly terraced by Lemma
\ref{lem:st_expr}.  Similarly, if there is an edge incident to distinct
vertices $u,v$ where $F_u$ is a parity constraint, contract that edge.
The \sig{} $F$ introduced by the contraction is a parity-\sig{} of
$F_v$, so this process does not increase $\sum_{v\in
  V}\Prat(F_v)$.  And again, $F$ is strictly terraced by Lemma
\ref{lem:st_expr}.  Repeating this process we end up with a circuit
with at most two vertices.  If there is only one vertex we can
appeal to the $k\leq 1$ case, and otherwise we are done by
Lemma~\ref{lem:pratrtfhelper}.

For $k>2$, contract any internal edge.  From
the $k\leq 2$ case we know that $\sum_{v\in V}\Prat(F_v)$ has not
increased.  This process does not change the \sig{} of $\phi$, and by
Lemma \ref{lem:st_expr} the constraint function introduced by the
contraction is strictly terraced.
\end{proof}

\begin{lemma}\label{lem:stbd}
Let $\phi$ be a closed circuit using strictly terraced constraints, and assume that $Z_0(\phi)>0$. Then
\begin{align}
\frac{Z_2(\phi)}{Z_0(\phi)} \leq \frac{1}{2}|E^\phi|^2 \max\left(1,\sum_{v\in V^\phi} \Prat(F^\phi_v)\right)^2.
\end{align}
\end{lemma}
\begin{proof}

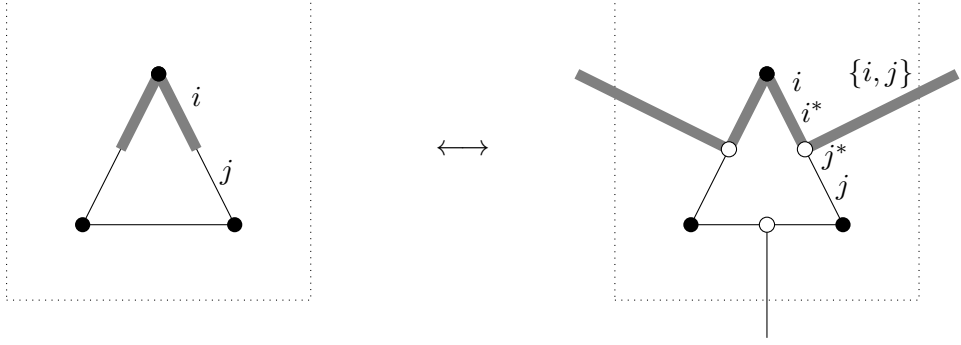
\begin{figure}
  \centering
\begin{tikzpicture}
  \begin{scope}
  \draw[dotted] (0,0) rectangle (4,4);
  \node at (2.5,2.7) {$i$};
  \node at (2.9,1.7) {$j$};
  \draw (2,3) -- (3,1) -- (1,1) -- (2,3);
  \draw[spun] (2,3) -- (2.5,2);
  \draw[spun] (2,3) -- (1.5,2);
  \foreach \x/\y in {2/3,3/1,1/1,2/3} \drawvert{\x}{\y};
  \end{scope}
  \node at (6,2) {$\longleftrightarrow$};  
  \begin{scope}[xshift=8cm]
  \node at (2.4,2.9) {$i$};
  \node at (2.6,2.5) {$i^*$};
  \node at (2.9,1.9) {$j^*$};
  \node at (3.0,1.5) {$j$};
  \node at (3.5,3.0) {$\{i,j\}$};
  \draw[dotted] (0,0) rectangle (4,4);
  \draw (2,3) -- (3,1) -- (1,1) -- (2,3);
  \draw[spun] (2,3) -- (2.5,2);
  \draw[spun] (2,3) -- (1.5,2);
  \draw[spun] (2.5,2) -- (4.5,3);
  \draw (2,1) -- (2,-0.5);
  \draw[spun] (1.5,2) -- (-0.5,3);
  \foreach \x/\y in {2/3,3/1,1/1,2/3} \drawvert{\x}{\y};
  \foreach \x/\y in {2.5/2,2/1,1.5/2} {
    \filldraw[fill=white] (\x,\y) circle (0.1cm);
    }
  \end{scope}
  
\end{tikzpicture}
\caption{ An illustration of the correspondence between
  arbitrary configurations in a closed circuit and assignments in a modified
  circuit.  Solid circles are arbitrary constraints, empty circles are
  copies of Even$_3$.  Thin black lines are incidences given the value
  $0$, thick grey lines are incidences given the value $1$.}\label{fig:evening}
\end{figure}

We will consider a circuit $\psi$ obtained by attaching $\sigEven_3$
to edges of $\phi$ as illustrated in Figure \ref{fig:evening}. In words:
let
$J^*$ be a disjoint copy of $J$, consisting of an element $i^*$ for
each $i\in J$.  Define $\psi$ to have incidences $E\cup J\cup J^*$,
external edges $E$, edges $\{i,i^*\}$ for each $i\in J$, vertex
set $V\cup E$, the same constraints at each $v\in V$, and
$F^{\psi}_{\{i,j\}}=\sigEven_{\{i^*,j^*,\{i,j\}\}}$ for all $\{i,j\}\in
E$.

$Z_k(\phi)$ is the sum of $\sem{\psi}(\vecx)$ over configurations
$\vecx$ of $E^\phi$ with $\sum_{e\in E^{\phi}}x_e=k$. By pinning,
$\Prat$ bounds the ratio $F(\vecx\oplus\eee i)/F(\vecx)$
between the weights of neighbouring
configurations of non-zero weight.
Letting $\veczero$ denote the all-zeros vector,
for all $i\neq j$ such that
$\sem{\psi}(\eee i)\neq 0$,
$$\sem{\psi}(\eee i+\eee j)\leq \Prat(\psi)\sem{\psi}(\eee i)\leq
\Prat(\psi)^2\sem{\psi}(\veczero).$$
If $\sem{\psi}(\eee i)= 0$ we have $\sem{\psi}(\eee i+\eee
j)=\sem{\psi}(\veczero)$ because $\sem{\psi}$ is strictly terraced.
Thus
$$Z_2(\phi) \leq Z_0(\phi) \binom{|E^\phi|}{2} \max(1,\Prat(\psi))^2.$$ The
result follows by applying Lemmas \ref{lem:pratex} and \ref{lem:pratbd}.
\end{proof}

\section{Proofs of Theorem \ref{thm:nappalg} and Theorem \ref{thm:nappexpr}}\label{sec:mainproofs}

\begin{reptheorem}{thm:nappalg}
$\napp$ has an FPRAS.
\end{reptheorem}
\begin{proof}
We are given a labelled graph, which is naturally a closed circuit $\phi$
using constraints of the form $\sigEven_J$, $\sigOdd_J$, and $\sigNAE_J$.

The decision problem, deciding whether $Z_0(\phi)>0$, can be solved in
polynomial time by Cornu\'ejols' algorithm for the general factor
problem \cite{cornuejols}. And degree-$1$ parity relations can be used
to fix edges to take a particular value. This means that $\napp$ is
self-reducible in the sense of \cite[Theorem 6.4]{JVV86}. So it suffices to give a
fully polynomial almost uniform sampler (FPAUS): an algorithm that,
when given an error parameter $0<\delta<1$ and an instance of $\napp$
corresponding to a closed circuit $\phi$ with $Z_0(\phi)>0$
using parity and not-all-equal relations, outputs an
assignment $\vecz$ satisfying
\[ \frac 1 2 \sum_{\vecx}|\Pr(\vecx=\vecz)-\wt_\phi(\vecx)/Z_0(\phi)| \leq \delta \]
in time polynomial in the size of the input and in $\log\delta^{-1}$.

We will use the near-assignments chain to sample from assignments of
$\phi$.  Define $F'\sigf J$ by
\begin{align*}
F'(\vecx)&=\prod_{v\in V}F_v(\vecx|_{J_v})\
\intertext{and}
F(\vecx)&=\begin{cases}F'(\vecx)& \sum_{i\in J}x_i\text{ is even}\\
0&\text{ otherwise.}
\end{cases}
\end{align*}

By Lemma \ref{lem:flipdecomp}, all the constraints of $\phi$ are
windable. By Lemma \ref{lem:windexpr}, $F'$ is windable.  By Lemma
\ref{lem:windchar}, $(F')_\oplus$ is even-windable.  But $F$ is a
pinning of $(F')_\oplus$ (setting the parity bit to zero).  A pinning
of an even-windable function is even-windable - this is immediate from
the characterisation in terms of 2-decompositions given in Section
\ref{sec:evenwind_examples}.

We will use the notation $\pi,\Omega,\Omega_0$ from Theorem
\ref{thm:namix}, for the near-assignment chain on the pair $(F,E^\phi)$.

Recall from Lemma \ref{lem:pratex} that $\Prat(\sigNAE_J)\leq 3$ and
$\Prat(\sigEven_J)=\Prat(\sigOdd_J)=0$, and that all these \sigs{} are
strictly terraced.  Let $R=3|V|^2|E|^2$; by Lemma \ref{lem:stbd}
we have $1/R\leq Z_0(\phi)/Z_2(\phi)\leq
Z_0(\phi)/(Z_0(\phi)+Z_2(\phi))=\pi(\Omega_0)$.

By Cornu\'ejols' algorithm, mentioned above, we get an assignment
$\vecy$ with $\sem{\phi}(\vecy)>0$ and in particular $\pi(\vecy)\geq 2^{-|E|}$.
Applying Theorem \ref{thm:namix}, by simulating the near-assignments
Markov chain of $(F,E)$ for $t\geq (2|E|)^4R^2(\log \frac
{2R}{\delta}+|E| \log 2)$ steps we can take a sample $\vecz$
from near-assignments of $\phi$ such that
\[ \frac 1 2 \sum_{\vecx\in\Omega}|\Pr(\vecx=\vecz)-\pi(\vecx)| \leq \delta/2R \]
Thus
\[ \frac 1 2 \sum_{\vecx\in\Omega_0}|\Pr(\vecx=\vecz|\vecz\in\Omega_0)-F(\vecx)/Z_0(\phi)| \leq \delta/2 \]

So we get an FPAUS by rejection sampling: run the simulation at least
$2R\log \frac 2 \delta$ times and return the first sample in
$\Omega_0$. The probability that this fails is small (at most
$(1-\frac 1 2 \pi(\Omega_0))^{2R\log \frac 2 \delta}\leq \delta/2$).
\end{proof}

\begin{reptheorem}{thm:nappexpr}
Let $\calF$ be the class of strictly terraced windable functions. Then
\begin{itemize}
  \item $\calF$ is closed under taking \sigs{} of connected circuits
  \item $\calF$ contains $\sigEven_k$, $\sigOdd_k$, and $\sigNAE_k$
    for all $k\geq 1$
  \item for all finite subsets $\calF'\subset\calF$ there is an FPRAS for $\Holant(\calF')$
\end{itemize}
\end{reptheorem}

\begin{proof}
The first statement is Lemma \ref{lem:windexpr}.  The second statement
is Lemma \ref{lem:flipdecomp}.

For the third statement, given $\calF'$, let
$\calF''=\calF'\cup\{\sigEven_1,\sigOdd_1\}$.  We can use $\sigEven_1$
and $\sigOdd_1$ to fix the value an edge takes, so $\Holant(\calF'')$
is self-reducible. (There is a minor difference from \cite{JVV86}: we
allow rational-valued functions. But this is not important.) For the
decision problem we can use Feder's algorithm for coindependent
relations \cite[Theorem 4]{feder}.  Otherwise the argument proceeds as
in the previous proof, taking $R$ to be $ |E|^2|V|^2
\max_{F\in\calF'}\Prat(F)$.  We find that $\Holant(\calF'')$, and
therefore $\Holant(\calF')$, has an FPRAS.
\end{proof}

\section{Matchings circuits}\label{sec:mgadg}

Define a \defn{matchings circuit} $G$ to be a graph fragment equipped with:
\begin{itemize}
\item a non-negative rational edge-weight $w(e)$ for each internal edge $e$
\item a non-negative rational fugacity $\lambda(v)$ for each vertex $v$
\end{itemize}
Note that in this definition the external edges are not given weights.

Let $\deg_F(v)$ denote the number of edges in $F$ incident to the
vertex $v$.  The weight of a set of edges $F\subseteq A\cup E$ is:
$$\wt_G(F)=
\begin{cases}
0&\text{ if $\deg_F(v)\geq 2$ for all vertices $v$}\\
\left(\prod_{\deg_F(v)=0}\lambda(v) \right)\left(\prod_{e\in F} w(e)\right)&\text{ otherwise.}
\end{cases}
$$

The \sig{} of $G$ is the function $\sem{G}\sigf A$
where $A$ is the set of external edges and
\[ \sem{G}(\vecx)=\sum_{\substack{F\subseteq E\\F\cap A=\{e\in A\mid x_e=1\}}}\wt_\phi(F). \]
As with circuits, if $F=\sem{G}$ we will say the $F$ \defn{has} the
matchings circuit $G$.

For all $w\geq 0$ define $\Edge^w\sigf 2$ by
\begin{align*}
\Edge^w(i,j)=\begin{pmatrix}1&0\\0&w\end{pmatrix}_{i,j}
\end{align*}
where the matrix rows and columns are indexed by $\{0,1\}$.
For all $\lambda\geq 0$ and all finite sets $J$ define $\Fugacity^\lambda_J\sigf J$ by
$$\Fugacity^{\lambda}_J(\vecx)=
\begin{cases}
\lambda&\text{ if $\sum_{i\in J}x_i=0$}\\
1&\text{ if $\sum_{i\in J}x_i=1$}\\
0&\text{ otherwise.}
\end{cases}$$

Given $G$, define a circuit by equipping each vertex $v$ with the
function $\Fugacity^{\lambda(v)}_{J_v}$, then subdividing each edge
$e$ and equipping the new vertex with the function $\Edge^{w(e)}$. The
circuit clearly has the same \sig{} as the matchings circuit.
So matchings circuits are just a special type of circuit.
We will use the same notation and terminology.

\subsection{Example}

\begin{proposition}\label{prop:nand}
For all finite sets $J$ define
$\relOR_J=\{\vecx\in\{0,1\}^J\mid \sum_{i\in J}x_i>0\}$.  Then
$\sigOR_J$ has a matchings circuit.
\end{proposition}
\begin{proof}
\begin{figure}
\begin{tikzpicture}
\draw[dotted] (-0.5,0) rectangle (10.5,4);

\foreach \i in {1,3,8} {
\begin{scope}[xshift=\i cm]
  \draw[dashed] (-0.2,0.5) rectangle (1.2,3.5);
  \draw (0,1) -- (1,1) -- (0.5,2) -- (0,1);
  \filldraw (0.5,3) circle (0.1cm);
  \filldraw (0,1) circle (0.1cm);
  \filldraw (1,1) circle (0.1cm);
  \draw (0.5,2) to [out=90+30,in=270-30] (0.5,3);
  \draw (0.5,2) to [out=90-30,in=270+30] (0.5,3);
  \draw (0.5,3) -- (0.5,4.5);
  \draw (1,1) to [out=270,in=270] (2,1);
  \filldraw[fill=white] (0.5,2) circle (0.1cm);
\end{scope}
}

  \draw (0,1) to [out=270,in=270] (1,1);
  \draw (7,1) to [out=270,in=270] (8,1);

  \draw (10,1) -- (10,2);
  \fill (0,1) circle (0.1cm);
  \fill (10,1) circle (0.1cm);
  \fill (10,2) circle (0.1cm);

  \draw node at (6,1.5) {$\dots$};
  \draw node at (6,4.2) {$\dots$};
\end{tikzpicture}
\caption{$2^{k-1}$OR$_k$ matchings circuit. Hollow circles are vertices with
  fugacity $1$. All other vertices have fugacity $0$, and all edges
  have edge-weight $1$.}\label{fig:orgadg}
\end{figure}
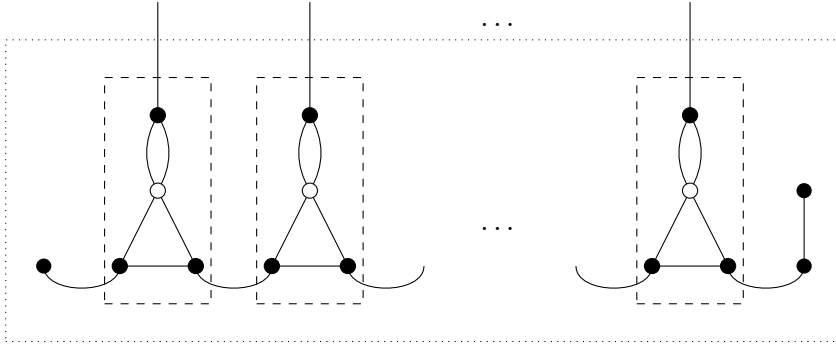

We may assume $J=\{1,\dots,k\}$.  The matchings circuit is
illustrated in Figure \ref{fig:orgadg}.

Define $F\sigf 3$ by
\begin{align*}
F(i,0,j)&=\begin{pmatrix}2&0\\0&2\end{pmatrix}_{i,j}\\
F(i,1,j)&=\begin{pmatrix}1&1\\1&1\end{pmatrix}_{i,j}
\end{align*}
(with rows and columns indexed from zero.)  Each of the
smaller boxes shown in Figure \ref{fig:orgadg} have the \sig{} $F$
(where external edges are numbered from left to right).

For all $x_1,\dots,x_k\in\{0,1\}$,
\begin{align*}
\sem{G}(x_1,\dots,x_k)
&=\sum_{y_1,\dots,y_{k-1}}F(1,x_1,y_1)F(y_1,x_2,y_2)\dots F(y_{k-1},x_k,0)\\
&=\left(\begin{pmatrix}2&0\\0&2\end{pmatrix}^{k-x_1+\dots-x_k}\begin{pmatrix}1&1\\1&1\end{pmatrix}^{x_1+\dots+x_k}\right)_{1,0}\\
&=2^{k-1}\sigOR_k(x_1,\dots,x_k).
\end{align*}

So the \sig{} of $G$ is $2^{k-1}\sigOR_k$.
To deal with the scalar multiple, add an isolated vertex with fugacity $1/2^{k-1}$.
\end{proof}

In particular, let $k\geq 1$ be odd.  Then $(\sigOR_k)_{\oplus}$ is a
copy of $\sigEvenOR_{k+1}$ where $\relEvenOR_{k+1}$ is defined as
$\relEven_{k+1}\cap\relOR_{k+1}$.  By Lemma \ref{lem:windchar},
$\sigEvenOR_{k+1}$ is even-windable.  Thus
$\sigEvenOR_{k+1}\overline{\sigEvenOR_{k+1}}=\sigEvenNAE_{k+1}$ has a
2-decomposition. This gives an alternate proof of Lemma
\ref{lem:evenwind_paritynae} which, via Lemma \ref{lem:flipdecomp},
shows that $\sigNAE_J$ is windable for all finite sets $J$. But this
argument does not seem to show that $\sigNAE_J$ has a matchings
circuit.

\subsection{Approximate counting}\label{sec:reductions}

Define

\prob{\nPM} {A simple graph $G$} {The number of perfect matchings in
  $G$}

The aim of this section is to establish Proposition \ref{prop:pmexpr},
that $\Holant(\calF)\APred\nPM$ for any finite set $\calF$ of \sigs{}
of matchings circuits, showing that matchings circuits are a natural
choice of circuit for $\nPM$. We will reduce via the following
problem.

\prob{$\FugacityWeightedPM$} {A closed matchings circuit $\phi$ where
  fugacities and edge-weights are given as ratios of non-negative
  integers specified in binary}  {$Z_0(\phi)$}

The fugacities and edge-weights can both be simulated using matchings circuits.
A similar reduction appears in \cite{Tutte}.

\begin{lemma}\label{lem:pmwtsim}
There is a polynomial-time algorithm which, given non-negative
integers $p,q$ specified in binary, outputs a matchings circuit
$G_{p,q}$ whose fugacities are all $0$ and whose edge-weights are all
$1$, and with two external edges such that
$$\sem{G_{p,q}}(i,j)=\begin{pmatrix}
p&0\\0&q\end{pmatrix}_{i,j}\qquad\text{ for all $i,j\in\{0,1\}$.}$$
where we consider the rows and columns of the matrix to be indexed
from zero.
\end{lemma}
\begin{proof}
See Figure \ref{fig:gpq}.
\begin{figure}
\begin{tikzpicture}
\draw[dotted] (0,-1.5) rectangle (10,1.5);
\filldraw (-0.5,0) --
      (1,0) circle (0.1cm)
      (4,0) circle (0.1cm) --
      (5,0) circle (0.1cm) --
      (6,0) circle (0.1cm)
      (8,0) circle (0.1cm) --
      (9,0) circle (0.1cm) --
      (10.5,0);
\draw (1,0) to[out=85,in=95] (4,0);
\draw (1,0) to[out=35,in=145] (4,0);
\draw (1,0) to[out=25,in=155] (4,0);
\node[anchor=north] at (0.7,-0.2) {$s=v_{3,1}$};
\node[anchor=north] at (2,-0.7) {$v_{3,2}$};
\node[anchor=north] at (3,-0.7) {$v_{3,3}$};
\node[anchor=north] at (4.3,-0.2) {$t=v_{3,4}$};

\draw (1,0) to[out=-15,in=145] (2,-0.5);
\draw (1,0) to[out=-45,in=175] (2,-0.5);
\filldraw (2,-0.5) circle(0.1cm) -- (3,-0.5) circle(0.1cm);
\draw (3,-0.5) to[out=35, in=195] (4,0);
\draw (3,-0.5) to[out=5, in=225] (4,0);

\draw (6,0) to[out=25,in=155] (8,0);
\draw (6,0) to[out=-25,in=-155] (8,0);
\end{tikzpicture}
\caption{$G_{7,2}$, with one path in the copy of $G_{7,1}$
  labelled. All fugacities are $0$, all edge-weights are $1$.}
\label{fig:gpq}
\end{figure}
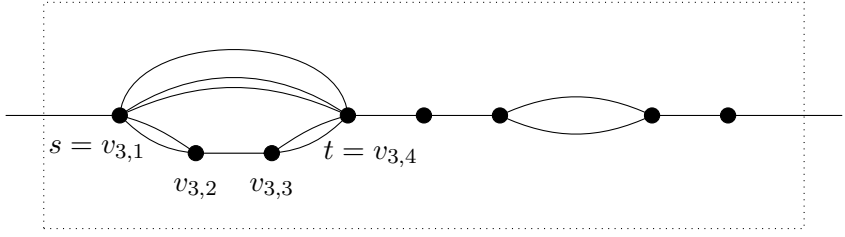

For all $p\geq 0$ there is a unique binary expansion
$p=2^{n_1}+\dots+2^{n_k}$, with $0\leq n_1<\dots<n_k$. Define
$G_{p,1}$ in the following way. Take two vertices $s$ and $t$, each
with one external edge. For each $1\leq i\leq k$, if $n_i=0$ add
an edge between $s$ and $t$, and otherwise add a path between $s$ and
$t$ of length $2n_i-1$, which we can denote
$s=v_{i,1},v_{i,2},\dots,v_{i,2n_i}=t$, and add a parallel edge in
the odd positions: between $v_{i,2j-1}$ and $v_{i,2j}$ for each
$1\leq j\leq n_i$.

There is a unique perfect matching of $G_{p,1}$ that includes the
external edges: it uses the edges in even position in each path,
$v_{i,2j}v_{i,2j+1}$ for all $1\leq i\leq k$ and all $1\leq j< n_i$.
The perfect matchings of $G_{p,1}$ that do not include the external
edges are determined by a choice of $1\leq i \leq k$ such that the
$i$'th path uses edges in odd positions, and a choice of edge in each
of the $n_i$ odd positions in this path; there are $2^{n_i}$ choices
for each $i$.  So $G_{p,1}$ has the correct \sig{}:
$\sem{G_{p,1}}(i,j)=\left(\begin{smallmatrix}
  p&0\\0&1\end{smallmatrix}\right)_{i,j}$.

Define $H$ to be the circuit consisting of one vertex with fugacity
zero, with two external edges.  For $q\neq 1$ define $G_{p,q}$ to be
serial composition of copies of $G_{p,1}$, $H$, $G_{q,1}$, and $H$,
that is, we identify the second external edge of the $i$'th circuit
with the first external edge of the $(i+1)$'th, for $i=1,2,3$.  The
\sig{} of $G_{p,q}$ is then given by the matrix
$$
\begin{pmatrix}p&0\\0&1\end{pmatrix}
\begin{pmatrix}0&1\\1&0\end{pmatrix}
\begin{pmatrix}q&0\\0&1\end{pmatrix}
\begin{pmatrix}0&1\\1&0\end{pmatrix}
=
\begin{pmatrix}p&0\\0&q\end{pmatrix}.
$$
\end{proof}

\begin{lemma}\label{lem:evenmatch}
Given a matchings circuit $G$ for a \sig{} $F$, we can efficiently
construct a matchings circuit $G'$ for $F_\oplus$ (defined in Section
\ref{sec:windable}) in which every vertex has fugacity zero.
Conversely, given a matchings circuit $G$ for $F_\oplus$, we can
efficiently construct a matchings circuit $G'$ for $F$.
\end{lemma}
\begin{proof}
\begin{figure}
\begin{tikzpicture}
\draw[dotted] (0.2,0) rectangle (9.8,5);

\foreach \i in {1,3,8} {
\begin{scope}[xshift=\i cm]
  \draw (0,1) -- (1,1) -- (0.5,2) -- (0,1);
  \filldraw (0.5,3) circle (0.1cm);
  \filldraw (0,1) circle (0.1cm);
  \filldraw (1,1) circle (0.1cm);
  \filldraw (0.5,2) circle (0.1cm);
  \draw (0.5,2) -- (0.5,3);
  \draw (-1,1) to [out=270,in=270] (0,1);
  \draw (0.5,3) --++ (-0.2,0.2);
  \draw (0.5,3) --++ (0,0.2);
  \draw (0.5,3) --++ (0.2,0.2);
\end{scope}
}
\draw node[anchor=east] at (1.5,2.6) {$\lambda(v_1)$};
\draw node[anchor=west] at (3.5,2.6) {$\lambda(v_2)$};
\draw node[anchor=west] at (8.5,2.6) {$\lambda(v_n)$};

  \draw node[anchor=east] at (1,1) {$b_1$};
  \draw node[anchor=west] at (2,1) {$c_1$};
  \draw node[anchor=west] at (1.5,2) {$a_1$};
  \draw node[anchor=west] at (1.5,3) {$v_1$};
  \draw node[anchor=east] at (3,1) {$b_2$};
  \draw node[anchor=west] at (4,1) {$c_2$};
  \draw node[anchor=west] at (3.5,2) {$a_2$};
  \draw node[anchor=west] at (3.5,3) {$v_2$};
  \draw node[anchor=east] at (8,1) {$b_n$};
  \draw node[anchor=west] at (9,1) {$c_n$};
  \draw node[anchor=west] at (8.5,2) {$a_n$};
  \draw node[anchor=east] at (8.5,3) {$v_n$};

  \draw (5,3.3) ellipse (4cm and 1cm);
\draw (2,3.8) -- (1,5.5);
\draw (3,3.8) -- (2,5.5);
\draw (7,3.8) -- (8,5.5);
  \draw (4,1) to [out=270,in=270] (5,1);
  \draw node at (5,3.3) {$F$};
  \draw node at (6,1.5) {$\dots$};
  \draw node at (6,2.5) {$\dots$};
  \draw node at (5,5.5) {$\dots$};
\end{tikzpicture}
\caption{Illustration of a matching circuit for $F_\oplus$ built from
  a matchings circuit for $F$, as described in Lemma
  \ref{lem:evenmatch}.}\label{fig:evenmatch}
\end{figure}
  For the first statement, pick an enumeration $v_1,\dots,v_n$ of the
  vertices of $G$.  Form $G'$ as follows.  For each $1\leq i\leq n$,
  add vertices $a_i,b_i,c_i$, edges $a_ib_i$,$a_ic_i$,$b_ic_i$ with
  edge-weight $1$, add an edge $v_ia_i$ with weight $\lambda(v_i)$,
  and if $i<n$ add an edge $c_ib_{i+1}$ with edge-weight $1$.
  Set all the fugacities to zero and add an external edge at $b_1$.  See Figure
  \ref{fig:evenmatch}.
  Consider a matching $M\subseteq E$ of $G$. We will argue that there
  is a unique way to extend $M$ to a perfect matching $M'$ of $G'$.
  
  Let $U=\{i\mid \deg_M(v_i)=0\}$ be the indices of unmatched
  vertices.  Let $M_1=\{a_iv_i\mid i\in U\}$. Note that the extension
  $M'$ must include $M_1$, and if $i\not\in U$ we cannot have $b_ic_i\in
  M'$. Consider the following subset $P$ of external and internal
  edges: the external edge at $b_1$, edges $b_ic_i$ for all $i\in U$,
  and the edges $b_ia_i$ and $a_ic_i$ for all $i\notin U$.  So $P$ is
  a path, except that at one endpoint, $P$ has an external edge $b_1$.
  Observe that there is a unique choice of perfect matching
  $M_2\subseteq P$ along this path: the odd-numbered edges
  starting from the end of $P$ not incident to the external edge $b_1$.
  (If $P$ has an odd number of vertices then we get $b_1\in M_2$, and otherwise $b_1\not\in M_2$.)
  Define $M'=M\cup M_1\cup
  M_2$. Any extension of $M$ to a perfect matching of $G'$ would have
  to include $M_1$, and hence $M_2$, and so the extension is unique.

  This gives a weight-preserving bijection between matchings $M$ of
  $G$ and perfect matchings $M'$ of $G'$.  Since $G'$ has an even
  number of vertices, the sets $M'$ must include an even number of
  external edges. Thus $\sem{G'}=F_\oplus$.

  The converse is easy: given a matchings circuit $G$ for $F_\oplus$,
  add a vertex of fugacity $1$ to the external edge $1$ to get a
  matchings circuit for $F$.
\end{proof}

\begin{lemma}\label{lem:fwpmp}
$\FugacityWeightedPM\APred\nPM$
\end{lemma}
\begin{proof}
Given a matchings circuit $G_1$ with no external edges, we will
construct a simple graph $G$ with $C\sem{G_1}$ perfect matchings
where $C$ is an easily computed positive integer.

By Lemma \ref{lem:evenmatch} we get a matchings circuit $G_2$ such that
$\sem{G_2}=\sem{G_1}_{\oplus}$. Deleting the external edge, we get a
circuit $G_3$ with $\sem{G_3}=\sem{G_1}$.  At each edge $e$ of $G_3$,
we have integers $p_e,q_e$ such that the weight of $e$ is $p_e/q_e$.
Insert a copy of the circuit $G_{p,q}$ given by Lemma
\ref{lem:pmwtsim}; this produces a circuit $G_4$ whose \sig{} is
$C\sem{G_3}$ where $C=\prod_{e\in E^{G_3}}q_e$, and where all
fugacities are $0$ and all edge-weights are $1$.  Forgetting the
fugacities and edge-weights we get a multigraph with $C\sem{G_3}$
perfect matchings.  To construct $G$, delete any loops and subdivide
each edge into a path of length $3$; this does not affect the number
of perfect matchings.
\end{proof}

\begin{repproposition}{prop:pmexpr}
If $\calF$ is a finite set of \sigs{} that have matchings circuits,
then $\Holant(\calF)\APred\nPM$.
\end{repproposition}
\begin{proof}
Pick a choice of matchings circuit $G_F$ for each $F\in \calF$.
Given an instance $\psi$ of $\Holant(\calF)$, for each vertex $v$ the
function $F_v$ is a copy of some $F\in \calF$; we can substitute
$G_F$ into $\psi$ at $v$. This process gives a matchings circuit
$G'$ with the same \sig{} as $\psi$.  We can then appeal to
Lemma \ref{lem:fwpmp}.
\end{proof}

\subsection{Expressive power}

\begin{lemma}\label{lem:pmwind}
The \sig{} of any matchings circuit is windable.
\end{lemma}
\begin{proof}
By Lemma \ref{lem:evenmatch} and Lemma \ref{lem:windchar} it suffices
to show that the \sig{} of any matchings circuit where all fugacities
are zero is even-windable.  For all $w\geq 0$, Lemma
\ref{lem:evenwind_small} implies that $\Edge^w$ is even-windable.  For
all $\lambda\geq 0$ and all finite sets $J$, consider a pinning
$G\sigf I$ of $\Fugacity^\lambda_J$.  If $(G\overline G)(\vecx)>0$ for
some $\vecx$ then $\sum_{i\in I} x_i$ and $\sum_{i\in I} (1-x_i)$ are at most $1$, so
$|I|\leq 2$.  Thus $G\overline G$ has a 2-decomposition as in
Lemma~\ref{lem:evenwind_small}.
\end{proof}

To give circuits for low-arity functions we will apply linear
programming duality in the form of Farkas' lemma. For a very short
proof of Farkas' lemma, as well as a statement explicitly allowing a
general ordered field, see \cite{Bartl}.  For all
$\vecx,\vecphi\in\mathbb{Q}^d$ denote the dot product
$x_1\phi_1+\dots+x_d\phi_d$ by $\vecx\cdot\vecphi$.

\begin{samepage}
\begin{lemma}\label{lem:farkas}
Let $\vecx_1,\dots,\vecx_k,\vecy\in \mathbb{Q}^d$.  The following are equivalent:
\begin{itemize}
  \item $\vecy=c_1\vecx_1+\dots+c_k\vecx_k$ for some $c_1,\dots,c_k\in\Qnonneg$
  \item $\vecy\cdot\vecphi\geq 0$ for all $\vecphi\in\mathbb{Q}^d$ that
satisfy $\vecx_1\cdot\vecphi, \dots,\vecx_k\cdot\vecphi\geq 0$
\end{itemize}
\end{lemma}
\end{samepage}

\begin{lemma}\label{lem:fiveclique}
Let $F\sigf 4$.  Assume that $F(\overline{\eee 1})$, $F(\overline{\eee
  2})$,$F(\overline{\eee 3})$,$F(\overline{\eee 4})$ are not all zero,
and that $F(x_1,x_2,x_3,x_4)=0$ whenever $x_1+x_2+x_3+x_4$ is
even, and that for all $x_1,x_2,x_3,x_4\in\{0,1\}$ we have
\begin{align*}
&&&F(x_1,x_2,x_3,x_4)F(1-x_1,1-x_2,1-x_3,1-x_4)\\
&\leq &&F(x_1,x_2,1-x_3,1-x_4)F(1-x_1,1-x_2,x_3,x_4)\\
&+&&F(x_1,1-x_2,x_3,1-x_4)F(1-x_1,x_2,1-x_3,x_4)\\
&+&&F(x_1,1-x_2,1-x_3,x_4)F(1-x_1,x_2,x_3,1-x_4).
\end{align*}
Then $F$ has a matchings circuit.
\end{lemma}
\begin{proof}

\begin{figure}
  \centering
  \begin{tikzpicture}[xshift=4cm]
    \draw[dotted] (-0.5,-1.5) rectangle (6.5,3);
    \node (v1) at (0,0) {};
    \node (v2) at (2,-1) {};
    \node (v3) at (4,-1) {};
    \node (v4) at (6,0) {};
    \node (u) at (3,2.5) {};
    \draw (0,0) -- (0,-2);
    \draw (2,-1) -- (2,-2);
    \draw (4,-1) -- (4,-2);
    \draw (6,0) -- (6,-2);
    \draw (u) -- node[anchor=east] {$F(\overline{\eee 1})$} (v1);
    \draw (u) -- node[anchor=east] {$F(\overline{\eee 2})$} (v2);
    \draw (u) -- node[anchor=west] {$F(\overline{\eee 3})$} (v3);
    \draw (u) -- node[anchor=west] {$F(\overline{\eee 4})$} (v4);
    \draw (v1) -- (v2) -- (v3) -- (v4) -- (v1);
    \draw (v1) -- (v3)  (v2) -- (v4);
    \begin{scope}[nodes={draw,fill=white}]
    \node[circle] at (v1) {$v_1$};
    \node[circle]  at (v2) {$v_2$};
    \node[circle] at (v3) {$v_3$};
    \node[circle] at (v4) {$v_4$};
    \node[circle] at (u) {$u$};
    \end{scope}
    \end{tikzpicture}
  \caption{A weighted clique. All fugacities are zero, and $w(uv_i)=F(\overline{\eee i})$ for all
    $i$. The other edge-weights are to be determined.}\label{fig:clique}
\end{figure}
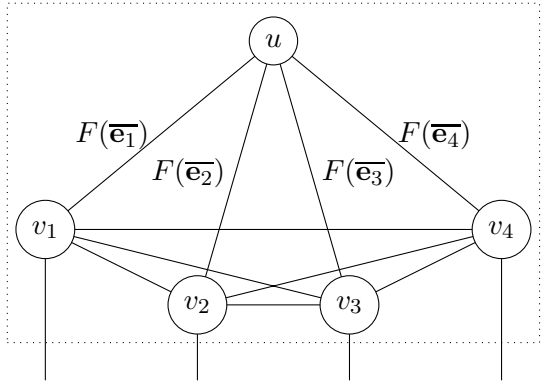

We will construct values $w(v_iv_j)\geq 0$ satisfying
\begin{align*}
F(\eee 1)&=F(\overline{\eee 2})w(v_3v_4)+F(\overline{\eee 3})w(v_4v_2)+F(\overline{\eee 4})w(v_2v_3)\\
F(\eee 2)&=F(\overline{\eee 3})w(v_4v_1)+F(\overline{\eee 4})w(v_1v_3)+F(\overline{\eee 1})w(v_3v_4)\\
F(\eee 3)&=F(\overline{\eee 4})w(v_1v_2)+F(\overline{\eee 1})w(v_2v_4)+F(\overline{\eee 2})w(v_4v_1)\\
F(\eee 4)&=F(\overline{\eee 1})w(v_2v_3)+F(\overline{\eee 2})w(v_3v_1)+F(\overline{\eee 3})w(v_1v_2)
\end{align*}
(and also $w(v_iv_j)=w(v_jv_i)$.)  This suffices because then $F=\sem{G}$
where $G$ is the weighted clique illustrated in Figure \ref{fig:clique}.
We need to show that the vector
$$\vecy=(F(\eee 1),F(\eee 2),F(\eee 3),F(\eee 4))$$ is a non-negative
linear combination of the vectors $\eee i F(\overline{\eee j})+\eee j
F(\overline{\eee i})$ with $1\leq i < j\leq 4$.  By Farkas' lemma (Lemma
\ref{lem:farkas}), it suffices to show that $\vecy\cdot\vecphi\geq 0$
for all $\vecphi\in\mathbb{Q}^4$ satisfying
\begin{align}\label{eq:phiij}
\phi_i F(\overline{\eee j})+\phi_j F(\overline{\eee i})\geq 0\qquad\text{ for all $1\leq i < j\leq 4$.}
\end{align}
If $\phi_1,\phi_2,\phi_3,\phi_4\geq 0$ we are done.
Otherwise $\phi_i<0$ for some $i$.
By assumption $F(\overline{\eee j})>0$ for some $j$.
If $j\neq i$, then \eqref{eq:phiij} implies that $\phi_j F(\overline{\eee i})$ is non-zero.
In any case $F(\overline{\eee i}) > 0$.
If $i=1$ then
\begin{align*}
F(\eee 1)F(\overline{\eee 1}) &\leq F(\eee 2)F(\overline{\eee 2}) + F(\eee 3)F(\overline{\eee 3}) + F(\eee 4)F(\overline{\eee 4})\\
-\phi_1 F(\eee 1)F(\overline{\eee 1}) &\leq -\phi_1 F(\eee 2)F(\overline{\eee 2}) -\phi_1 F(\eee 3)F(\overline{\eee 3})-\phi_1 F(\eee 4)F(\overline{\eee 4})\\
-\phi_1 F(\eee 1)F(\overline{\eee 1}) &\leq \phi_2 F(\eee 2)F(\overline{\eee 1}) + \phi_3 F(\eee 3)F(\overline{\eee 1}) + \phi_4 F(\eee 4)F(\overline{\eee 1})\\
-\phi_1 F(\eee 1) &\leq \phi_2 F(\eee 2) + \phi_3 F(\eee 3) + \phi_4 F(\eee 4)
\end{align*}  Therefore $\vecy\cdot\vecphi\geq 0$.
By symmetry the other cases, $i\neq 1$, are similar.
\end{proof}

\begin{reptheorem}{thm:arity3}
Let $F\sigf 3$. The following are equivalent:
\begin{enumerate}
\item $F$ is windable
\item For all $x_1,x_2,x_3\in\{0,1\}$ we have
\begin{align*}
&&&F(x_1,x_2,x_3)F(1-x_1,1-x_2,1-x_3)\\
&\leq &&F(x_1,x_2,1-x_3)F(1-x_1,1-x_2,x_3)\\
&+&&F(x_1,1-x_2,x_3)F(1-x_1,x_2,1-x_3)\\
&+&&F(x_1,1-x_2,1-x_3)F(1-x_1,x_2,x_3)
\end{align*}
\item $F$ has a matchings circuit
\end{enumerate}
\end{reptheorem}

\begin{proof}
For notational convenience, in the following argument we will use a
particular copy of $F_\oplus$. For all $x_1,x_2,x_3,x_4\in\{0,1\}$ define
$$F'(x_1,x_2,x_3,x_4)=\begin{cases}
F(x_1,x_2,x_3)&\text{ if $x_1+x_2+x_3+x_4$ is even}\\
0&\text{ otherwise.}
  \end{cases}
  $$
  
($1\implies 2$) Let $B$ be a witness that $F'$ is even-windable
  (using Lemma \ref{lem:windchar}).
Let $x_1,x_2,x_3\in\{0,1\}$.
Let $c\in\{0,1\}$ be the
unique value such that $x_1+x_2+x_3+c$ is even. Then
$(x_1,x_2,x_3,c)\oplus (1-x_1,1-x_2,1-x_3,1-c)=(1,1,1,1)$.
Note that
$$\Match{(1,1,1,1)}=\{\{\{1,2\},\{3,4\}\},\{\{1,3\},\{2,4\}\},\{\{1,4\},\{2,3\}\}\}.$$
We have
\begin{align*}
&F(x_1,x_2,x_3)F(1-x_1,1-x_2,1-x_3)\\
&=F'(x_1,x_2,x_3,c)F'(1-x_1,1-x_2,1-x_3,1-c)\\
&=B((x_1,x_2,x_3,c),(1-x_1,1-x_2,1-x_3,1-c),\{\{1,2\},\{3,4\}\})\\
&\hspace{.5in}+B((x_1,x_2,x_3,c),(1-x_1,1-x_2,1-x_3,1-c),\{\{1,3\},\{2,4\}\})\\
&\hspace{.5in}+B((x_1,x_2,x_3,c),(1-x_1,1-x_2,1-x_3,1-c),\{\{1,4\},\{2,3\}\})\\
&=B((x_1,x_2,1-x_3,1-c),(1-x_1,1-x_2,x_3,c),\{\{1,2\},\{3,4\}\})\\
&\hspace{.5in}+B((x_1,1-x_2,x_3,1-c),(1-x_1,x_2,1-x_3,c),\{\{1,3\},\{2,4\}\})\\
&\hspace{.5in}+B((x_1,1-x_2,1-x_3,c),(1-x_1,x_2,x_3,1-c),\{\{1,4\},\{2,3\}\})\\
&\leq F'(x_1,x_2,1-x_3,1-c)F'(1-x_1,1-x_2,x_3,c)\\
&\hspace{.5in}+F'(x_1,1-x_2,x_3,1-c)F'(1-x_1,x_2,1-x_3,c)\\
&\hspace{.5in}+F'(x_1,1-x_2,1-x_3,c)F'(1-x_1,x_2,x_3,1-c)\\
&= F(x_1,x_2,1-x_3)F(1-x_1,1-x_2,x_3)\\
&\hspace{.5in}+F(x_1,1-x_2,x_3)F(1-x_1,x_2,1-x_3)\\
&\hspace{.5in}+F(x_1,1-x_2,1-x_3)F(1-x_1,x_2,x_3)
\end{align*}

($2\implies 3$) We can assume that $F$ is not identically zero
(otherwise, take two vertices of fugacity $0$, and attach four
outgoing edges to one of them - the isolated vertex can never be
matched).
Pick $\vecx\in\{0,1\}^4$ with
$F'(x_1,x_2,x_3,x_4)>0$. Lemma \ref{lem:fiveclique} implies that the
flip $F''$ of $F'$ by $\vecx\oplus (1,1,1,0)$ has a matchings circuit.
By subdividing the $i$'th outgoing edge for each $i$ with $x_i=1$, we
get a matchings circuit for $F'$.  By Lemma \ref{lem:evenmatch} we get
a matchings circuit whose \sig{} is $F$.

($3\implies 1$) Lemma \ref{lem:pmwind}.
\end{proof}

In particular by Theorem \ref{thm:arity3} and Proposition \ref{prop:pmexpr}, $\Holant(\{\mathbf R\})\APred\nPM$ where
$$R=\{(0,0,0),(1,0,0),(0,1,0),(1,0,1),(0,1,1)\}.$$

\bibliographystyle{plain} \bibliography{main}

\end{document}